\documentclass[A_4paper,11pt, final]{article}
\usepackage[utf8]{inputenc}
\usepackage{jheppub}
\usepackage{amsmath}
\usepackage{float}
\usepackage{tabularx}

\newcommand*\diff{\mathrm{d}}

\usepackage{xspace}
\newcommand*{\ie}{i.e., }
\newcommand*{\eg}{e.g., }

\newcommand*{\eq}{eq.\@\xspace}
\newcommand*{\eqs}{eqs.\@\xspace}
\newcommand*{\cf}{cf.\@\xspace}
\newcommand*{\etc}{etc.\@\xspace}

\title{Coupling Metric-Affine Gravity\newline
	to the Standard Model and Dark Matter Fermions}

\author[a]{Claire Rigouzzo,}
\author[b]{Sebastian Zell}

\affiliation[a]{Laboratory for Theoretical Particle Physics and Cosmology,\\
	King's College London, London, United Kingdom}
\affiliation[b]{Centre for Cosmology, Particle Physics and Phenomenology -- CP3,
	Universit\'e catholique de Louvain, B-1348 Louvain-la-Neuve, Belgium}
\emailAdd{claire.rigouzzo@kcl.ac.uk}
\emailAdd{sebastian.zell@uclouvain.be}

\abstract{General Relativity (GR) exists in different formulations, which are equivalent in pure gravity. Once matter is included, however, observable predictions generically depend on the version of GR. In order to quantify the resulting ambiguity, we employ metric-affine gravity, which encompasses as special cases the metric, Palatini, Einstein-Cartan and Weyl formulations. We first discuss the interaction of fermions with torsion and non-metricity, also commenting on projective symmetry. With a view towards the Standard Model, we then construct a generic model of (complex) scalar, fermionic and gauge fields coupled to GR and derive an equivalent metric theory, which features numerous new interaction terms. As a first observable consequence, we point out that a gravitational mechanism for producing dark matter in the form of singlet fermions can be used to distinguish between metric gravity and other formulations of GR.}

\date{}

\makeatletter
\gdef\@fpheader{\phantom{text}}
\makeatother

\begin{document}
	
	\maketitle
	\pagebreak
	\section{Introduction}
	\label{sec:intro}
	
	\paragraph{Overall goal.}
General Relativity (GR) has been confirmed by countless experiments, such as the recent breakthrough discovery of gravitational waves \cite{LIGO}. Arguably, however, this theory is far from being fully understood. An important question already arises on the classical level: Which formulation of GR should one use? All options are fully equivalent in pure gravity, provided that the action is chosen to be sufficiently simple. In other words, GR cannot distinguish between its different versions, which puts all of them on the same footing. \textit{Therefore, the formulations of GR do not represent a modification of gravity. Instead, they have to be regarded as unavoidable ambiguity of GR.}
	
	Once matter is included, the equivalence is broken and the various versions of GR generically lead to distinct predictions. On the one hand, this is a very undesirable feature. In principle, it becomes impossible to make unique predictions in all theories of matter coupled to GR. One the other hand, an opportunity arises for using measurements to derive information about GR: One can constrain the choice of formulation by observations. In the present paper, we shall contribute to this goal. Throughout, our approach will be minimal, \ie we shall solely consider those ambiguities that are forced upon us by gravity. Consequently, we only employ models that are equivalent to metric GR in the absence of matter.
	
	\paragraph{Different formulations.} Gravity is described by means of the geometry of spacetime, determined by the metric $g_{\mu \nu}$ and the affine connection $\Gamma^\alpha_{\ \mu \nu}$. In the original version of GR \cite{Einstein:1915}, two constraints were imposed on $\Gamma^\alpha_{\ \mu \nu}$, namely metric-compatibility, $\nabla_\alpha g_{\mu \nu=0}$, and the absence of torsion, $\Gamma^\alpha_{\ \mu \nu}= \Gamma^\alpha_{\ \nu \mu}$. These two requirements uniquely determine $\Gamma^\alpha_{\ \mu \nu}$ to be the Levi-Civita connection $\mathring{\Gamma}^\alpha_{~\beta \gamma}(g_{\mu\nu})$, which is a function of the metric. This is the metric formulation of GR, where  $g_{\mu \nu}$ is the only fundamental field and geometry is Riemannian, \ie fully determined by the curvature of spacetime. 
	
	It was soon realized that the above assumptions can be relaxed and that the affine connection and the metric can constitute independent fundamental variables \cite{Weyl:1918, Palatini:1919,Weyl:1922, Eddington:1923, Cartan:1922, Cartan:1923, Cartan:1924, Cartan:1925,Einstein:1925, Einstein:1928, Einstein:19282}.\footnote
	{Translations of \cite{Palatini:1919}, \cite{Cartan:1922} and \cite{Einstein:1925,Einstein:1928, Einstein:19282} are provided in \cite{Hojman:1980}, \cite{Kerlick:1980} and  \cite{Unzicker2005}, respectively, and we refer to \cite{Ferraris:1981} for a historical discussion.}
	Then deviations from Riemannian geometry arise, encoded in torsion $T^\alpha_{\ \beta \gamma} =\frac{1}{2}(\Gamma^\alpha_{\ \beta \gamma} - \Gamma^\alpha_{\ \gamma\beta })$, the non-closure of infinitesimal parallelograms \cite{Cartan:1922,Cartan:1923,Cartan:1924,Cartan:1925}, as well as non-metricity, causing a change of vector length during parallel transport \cite{Weyl:1918, Weyl:1922,Eddington:1923}. If only torsion is included in addition to curvature, this leads to the Einstein-Cartan (EC) formulation of GR \cite{Cartan:1922,Cartan:1923,Cartan:1924,Cartan:1925,Einstein:1925, Einstein:1928,Einstein:19282}, whereas a theory that exclusively features non-metricity on top of curvature can be regarded as Weyl formulation of GR.\footnote
	{This is different from gravitational models that exhibit Weyl-invariance (see \eg \cite{Kallosh:2013hoa,Kallosh:2013daa,Karananas:2015ioa,Oda:2018zth,Barnaveli:2018dxo, Ghilencea:2018thl,Edery:2019txq,Ferreira:2019zzx,Lin:2020phk,Hobson:2020doi,Tang:2020ovf,Karananas:2021gco,Olmo:2022ops,Sauro:2022chz,Aoki:2022csb,Sauro:2022hoh,Nakayama:2022qbs,Lalak:2022wyu,Paci:2023twc}).}
	Ultimately, including all three geometric properties of gravity -- curvature, torsion and non-metricity -- corresponds to a general metric-affine version of GR \cite{Hehl:1976kt, Hehl:1976kv, Hehl:1976my, Hehl:1977fj}. This formulation stands out since no assumptions are made on the nature of the affine connection.\footnote
	{Reviews of Einstein-Cartan/metric-affine gravity are provided in \cite{Hehl:1980,Hehl:1994ue,Blagojevic:2002, Blagojevic:2003cg, Obukhov:2018bmf} and an overview of various versions of GR can be found in \cite{Heisenberg:2018vsk, BeltranJimenez:2019esp, Rigouzzo:2022yan}, where also other options are discussed such as teleparallel theories, in which curvature is assumed to vanish \cite{Einstein:1928, Einstein:19282, Moller:1961, Pellegrini:1963, Hayashi:1967se,Cho:1975dh, Hayashi:1979qx, Nester:1998mp, BeltranJimenez:2019odq}, as well as purely affine models, where the only dynamical field is the affine connection \cite{Eddington:1923,Einstein:1925,Schroedinger:1950, Kijowski:1978}.}

	Remarkably, even though those formulations may look drastically different, they are fully equivalent for sufficiently simple theories. If for instance no matter is coupled to GR and gravity is described by the Einstein-Hilbert action $\int \diff^4 x \sqrt{-g}\, R$, where $R$ is the Ricci scalar, then torsion and/or non-metricity vanish dynamically and we recover the Levi-Civita connection $\mathring{\Gamma}^\alpha_{~\beta \gamma}$ \cite{Dadhich:2012htv,Bernal:2016lhq}.\footnote
	{If -- possibly in the presence of matter -- the purely gravitational part of a theory only consists of $R$, we shall use the term Palatini formulation. Such a situation can be regarded as special case of the Einstein-Cartan, Weyl or metric-affine versions (see discussion in \cite{Rigouzzo:2022yan}).}
	Because of this equivalence, the different formulations represent an inherent ambiguity of GR. 
Various arguments have been put forward for preferring one formulation over the other. As an important example, EC gravity can be derived by gauging the Poincar\'e group \cite{Utiyama:1956sy,Kibble:1961ba,Sciama:1962}, which puts gravity on the same footing as the other forces of the Standard Model (SM). What speaks in favor of metric-affine gravity is the fact that it avoids a priori assumptions about the connection and encompasses the metric, Palatini and EC versions as special cases. Having said that, there is no irrefutable argument to single out any of the formulations and we have to regard all of them as intrinsic part of GR. 

	\paragraph{Breaking the equivalence.}
	There are two ways to break the equivalence between the various versions of GR. The first one already exists in pure gravity and utilizes curvature-squared terms, which generically introduce additional propagating degrees of freedom that are often plagued by inconsistencies \cite{Stelle:1977ry,Neville:1978bk,Neville:1979rb,Sezgin:1979zf,Hayashi:1979wj, Hayashi:1980qp}.
	Even though certain models are healthy at the linear \cite{Sezgin:1981xs,Kuhfuss:1986rb,Nair:2008yh,Nikiforova:2009qr,Karananas:2014pxa,Karananas:2016ltn,Obukhov:2017pxa,Blagojevic:2017ssv,Blagojevic:2018dpz,Lin:2018awc,BeltranJimenez:2019acz,Aoki:2019rvi,Jimenez:2019qjc,Lin:2019ugq, Percacci:2019hxn,BeltranJimenez:2020sqf,Marzo:2021esg,Marzo:2021iok,Baldazzi:2021kaf,Pradisi:2022nmh,Annala:2022gtl} or even full non-linear level \cite{Yo:1999ex,Yo:2001sy,BeltranJimenez:2014iie,Iosifidis:2018diy,Iosifidis:2018zwo,Helpin:2019vrv,OrejuelaGarcia:2020viw,Ghilencea:2020piz,BeltranJimenez:2020sih,Xu:2020yeg,Yang:2021fjy,Quiros:2021eju,Quiros:2022uns,Yang:2022icz,Burikham:2023bil,Haghani:2023nrm,Barker:2023fem}, we shall not consider them since the new particles cause a deviation from the metric version of GR already in the absence of matter.
	
	The second way to break the equivalence between the different formulations of GR is through the interaction with matter. In certain cases, \eg for fermions, this happens even when the coupling to gravity is minimal \cite{Kibble:1961ba, Rodichev:1961}. In other situations, a difference between the various versions of GR only arises for non-minimal interactions \cite{Hojman:19802, Nelson:1980ph,Nieh:1981ww, Percacci:1990wy, Castellani:1991et,Hehl:1994ue, Holst:1995pc,Obukhov:1996pf, Obukhov:1997zd,Shapiro:2001rz, Freidel:2005sn, Alexandrov:2008iy,Diakonov:2011fs, Magueijo:2012ug, Pagani:2015ema,Rasanen:2018ihz, Shimada:2018lnm}. These effects of a coupling to geometry can be mapped to an equivalent theory in the metric formulation with a specific set of operators of higher mass dimension in the matter sector. Thus, the different formulations of GR become distinguishable at sufficiently high energies. It is important to note, however, that no additional dynamical particles emerge from torsion and non-metricity, \ie the only propagating gravitational degree of freedom is still the massless spin $2$ graviton. 
	
	In EC gravity, the concrete form of higher-dimensional operators has been worked out in various models \cite{Kibble:1961ba,Nelson:1980ph, Castellani:1991et, Perez:2005pm, Freidel:2005sn,Alexandrov:2008iy,Taveras:2008yf, Torres-Gomez:2008hac, Calcagni:2009xz,Mercuri:2009zi, Diakonov:2011fs, Magueijo:2012ug,Langvik:2020nrs, Shaposhnikov:2020frq,Karananas:2021zkl}. So far, \cite{Karananas:2021zkl} represents the most complete study since it includes all fields of the SM and encompasses as special cases all previously mentioned works \cite{Kibble:1961ba,Nelson:1980ph, Castellani:1991et, Perez:2005pm, Freidel:2005sn,Alexandrov:2008iy,Taveras:2008yf, Torres-Gomez:2008hac, Calcagni:2009xz,Mercuri:2009zi, Diakonov:2011fs, Magueijo:2012ug,Langvik:2020nrs, Shaposhnikov:2020frq}. Moreover, systematic criteria were proposed in \cite{Karananas:2021zkl} for constructing an action of matter coupled to GR, with the goal of ensuring equivalence to the metric formulation in pure gravity while making as few assumptions as possible. Metric-affine scenarios were studied in \cite{Percacci:1990wy, Obukhov:1996pf, Obukhov:1997zd,Pagani:2015ema,Rasanen:2018ihz,Shimada:2018lnm,Iosifidis:2021bad,Rigouzzo:2022yan,Rasanen:2022ijc}, where so far an explicit coupling to the whole SM has not been considered.

	\paragraph{Phenomenological implications.}
In the absence of conclusive conceptual arguments to distinguish between the formulations of GR, it is crucial to derive from them observables predictions. In this way, one can use measurements to reduce the ambiguity that results from the different versions of GR. Phenomenological and cosmological implications of EC (or Palatini) gravity have been analyzed in \cite{Stoeger:1979,Lammerzahl:1997wk,Freidel:2005sn,Bauer:2008zj,Shie:2008ms, Taveras:2008yf, Torres-Gomez:2008hac,Chen:2009at,Baekler:2010fr,Poplawski:2011xf, Diakonov:2011fs, Khriplovich:2012xg, Magueijo:2012ug, Khriplovich:2013tqa,Markkanen:2017tun,Carrilho:2018ffi, Kranas:2018jdc,Rasanen:2018fom,Rubio:2019ypq,Zhang:2019mhd,Zhang:2019xek,Saridakis:2019qwt,Barman:2019mlj, Shaposhnikov:2020geh,Aoki:2020zqm,Karananas:2020qkp,Langvik:2020nrs,Shaposhnikov:2020gts,Shaposhnikov:2020aen,Kubota:2020ehu,Enckell:2020lvn,Iosifidis:2021iuw,Racioppi:2021ynx,Cheong:2021kyc, Benisty:2021sul,Piani:2022gon,Dux:2022kuk,Klaric:2022qly,Yin:2022fgo,Rasanen:2022ijc,Gialamas:2022gxv,Piani:2023aof}, among others (see \cite{Gialamas:2023flv} for a recent review focusing on inflation), and a list of corresponding studies in metric-affine gravity includes \cite{Obukhov:1997zd,Minkevich:1998cv,Puetzfeld:2001hk,Babourova:2002fn,Puetzfeld:2004yg,Latorre:2017uve,Rasanen:2018ihz,Shimada:2018lnm, Iosifidis:2020zzp,Mikura:2020qhc, Iosifidis:2020upr,Mikura:2021ldx,Iosifidis:2021kqo, Iosifidis:2021fnq,Gialamas:2022xtt,Iosifidis:2022xvp,Arai:2022ilw,Iosifidis:2023but,Iosifidis:2023pvz,Boehmer:2023fyl,Gialamas:2023emn}, where we note that additional propagating degrees of freedom arise from gravity in some of the above mentioned works.

Of particular importance for the present paper is a possible implication of the formulations of GR for dark matter. Among the proposed candidates, singlet fermions are especially well motivated since they can -- in the form of right-handed neutrinos -- additionally provide an explanation for the observed neutrino masses. For a long time, however, it was difficult to implement the production of fermionic dark matter in the early Universe while obeying observations constraints \cite{Dodelson:1993je,Shi:1998km,Shaposhnikov:2008pf,Canetti:2012vf,Canetti:2012kh,Ghiglieri:2020ulj} (see \cite{Boyarsky:2018tvu} for a review). In \cite{Shaposhnikov:2020aen}, it was demonstrated that EC gravity features a new and natural mechanism for generating singlet fermions.\footnote
{Dark matter production from dynamical torsion was discussed in \cite{Barman:2019mlj}.}
It can produce the observed abundance of dark matter in a wide range of fermion masses and moreover leads to a characteristic momentum distribution of dark matter, which can serve to probe this proposal. So far, however, it has remained unclear if a gravitational production of singlet fermions is possible in other formulations of GR apart from EC gravity. 
	
	\paragraph{Present work.}
In the present paper, we pursue three goals:
	\begin{enumerate}
		
		\item We will couple GR to a matter sector that includes a complex scalar field, fermions and vector gauge bosons and therefore contains the SM. To this end, we employ the metric-affine formulation that unifies the parameter space of the metric, Palatini, EC and Weyl versions. As in \cite{Rigouzzo:2022yan}, we will derive an equivalent metric theory. Our result encompasses the models \cite{Kibble:1961ba,Nelson:1980ph, Castellani:1991et, Perez:2005pm, Freidel:2005sn,Alexandrov:2008iy,Taveras:2008yf, Torres-Gomez:2008hac, Calcagni:2009xz,Mercuri:2009zi, Diakonov:2011fs, Magueijo:2012ug,Rasanen:2018ihz,Shimada:2018lnm,Langvik:2020nrs, Shaposhnikov:2020frq,Karananas:2021zkl,Rigouzzo:2022yan} as special cases. We emphasize that our study will be classical throughout, apart from very brief comments on quantum effects.
		
		\item Along the way, we shall discuss the coupling of fermions to torsion and non-metricity. Moreover, we will develop further the criteria of \cite{Karananas:2021zkl} for coupling matter to GR.
		\item Finally, we will show that a gravitational production of fermionic dark matter is not only possible in EC gravity \cite{Shaposhnikov:2020aen}, but it also occurs naturally in the metric-affine (as well as Weyl) formulations. Therefore, this mechanism can be viewed as a generic prediction of GR outside the metric version. Also observable predictions are largely insensitive to the different formulations of GR, as long as they feature an independent affine connection.
	\end{enumerate}

	The paper is organized as follows. After a brief review of the geometrical aspects of classical gravity, we discuss in section \ref{sec:Dev-Riemannian} the coupling of GR to matter fields, in particular fermions, and present the refined selection criteria for coupling matter to gravity. In section \ref{sec:Gen-Action}, the most general action satisfying the selection rules is constructed. We show that the presence of torsion and non-metricity leads to specific geometric-induced interaction terms for the fermion and scalar field whereas no such contributions emerge for the gauge fields. Finally, we study in section \ref{sec:darkmatter} the phenomenology of the new higher-dimensional operators for fermions and show that the gravitational production of dark matter proposed in \cite{Shaposhnikov:2020aen} can also be naturally realized in metric-affine gravity. We conclude in section \ref{sec:conclusion}, furthermore giving an outlook to implications for quantum gravity. In appendix \ref{app:localLorentz}, we discuss local Lorentz symmetry as well as the covariant derivative of fermions and appendix  \ref{app:criteria} contains details about the selection rules for coupling of gravity to different matter fields.
	 Appendix \ref{app_details_computations} is devoted to details about the computation of section \ref{sec:Gen-Action} and we also show that results of \cite{Karananas:2021zkl, Rigouzzo:2022yan} are reproduced in specific limits.

	\paragraph{Conventions.} Greek letters denote spacetime indices and Latin letters are reserved for Lorentz indices. Both the spacetime metric $g_{\mu\nu}$ and the Minkowski metric $\eta_{\alpha\beta}$ have signature $(-1,+1,+1,+1)$. Square brackets denote antisymmetrization, $T_{[\mu \nu]} \equiv \frac{1}{2} (T_{\mu\nu} - T_{\nu\mu})$, and round brackets indicate symmetrization, $T_{(\mu \nu)} \equiv \frac{1}{2} (T_{\mu\nu} + T_{\nu\mu})$. For the gamma matrices we use the convention
	\begin{equation} \label{gammaConvention}
		\left\{\gamma_A, \gamma_B \right\} = - 2 \eta_{AB} \;, \qquad \gamma_5 = -i \gamma^0 \gamma^1 \gamma^2 \gamma^3 = i \gamma_0 \gamma_1 \gamma_2 \gamma_3 \;,
	\end{equation}   
and for the Levi-Civita tensor we take: 
\begin{equation} \label{epsilonConvention}
\epsilon_{0123}=1=-\epsilon^{0123} \;.
\end{equation}   
	The covariant derivative of a vector $A^\nu$ is defined as
	\begin{equation} \label{covariantDerivative}
		\nabla_\mu A^\nu= \partial_\mu A^\nu + \Gamma^\nu_{~\mu \alpha}A^\alpha \;,
	\end{equation}
	\ie the summation is done on the last index of the Christoffel symbol. Finally, we work in natural units $M_P=\hbar=c=1$, where $M_P$ is the reduced Planck mass. 
	
	\section{The geometry of gravity}
	\label{sec:Dev-Riemannian}
	
	\subsection{Review of curvature, torsion and non-metricity}
	We shall give a brief overview of gravitational geometry and refer the reader to \cite{Heisenberg:2018vsk,Rigouzzo:2022yan} for more details. Our starting point is a differentiable manifold equipped with a metric $g_{\mu\nu}$ and a independent connection $\Gamma^\alpha_{~\beta \gamma}$, where the latter determines the Riemann tensor :
	\begin{equation} \label{riemannTensor}
		R_{~\sigma \mu \nu}^{\rho}=\partial_{\mu} \Gamma_{~\nu \sigma}^{\rho}-\partial_{\nu} \Gamma_{~\mu \sigma}^{\rho}+\Gamma_{~\mu \lambda}^{\rho} \Gamma_{~\nu \sigma}^{\lambda}-\Gamma_{~\nu \lambda}^{\rho} \Gamma_{\mu \sigma}^{\lambda} \;.
	\end{equation}
	Then the curvature of spacetime reads 
	\begin{equation} \label{RicciScalar}
		R = g^{\sigma \nu} R_{~\sigma \rho \nu}^{\rho} \;,
	\end{equation}
	corresponding to the rotation of vectors along infinitesimal closed curves \cite{Schutz:1980,Carroll:2004,Mielke:2017nwt}. If we were to impose that the connection is symmetric in the lower indices, $\Gamma^\alpha_{~\beta \gamma} = \Gamma^\alpha_{~\gamma \beta}$, and metric compatible, $\mathring{\nabla}_\mu g_{\alpha \beta}=0$, this would uniquely lead to the Levi-Civita connection $\mathring{\Gamma}^\alpha_{~\beta \gamma}$, which is a function of the metric:
	\begin{equation}
		\mathring{\Gamma}^\alpha_{~\beta \gamma}=\frac{1}{2}g^{\alpha \mu}(\partial_\beta g_{\mu \gamma}+\partial_\gamma g_{\mu \beta}-\partial_\mu g_{\beta \gamma}) \;.
		\label{levicivita}
	\end{equation}
	If assumptions about the connection are dropped, however, then two additional geometric properties arise in addition to curvature $R$.
	The first one is torsion,
	\begin{equation} \label{torsionDefinition}
		T^\alpha_{~\beta \gamma} \equiv \Gamma^\alpha_{~\beta \gamma}-\Gamma^\alpha_{~\gamma \beta} \;,
	\end{equation}
	and corresponds to the non-closure of infinitesimal parallelograms. Secondly, we have non-metricity,
	\begin{equation} \label{nonmetricityDefinition}
		Q_{\gamma \alpha \beta} \equiv \nabla_{\gamma}g_{\alpha \beta} \;,
	\end{equation}
	which leads to the non-conservation of vector norms under parallel transport.	
	
	We can uniquely decompose the full connection $\Gamma^\alpha_{~\beta \gamma}$ into the Levi-Civita part $\mathring{\Gamma}^\alpha_{~\beta \gamma}$ and non-Riemannian contributions:
	\begin{equation}
		\Gamma^\gamma_{~\alpha \beta}= \mathring{\Gamma}^\gamma_{~\alpha \beta}(g)+J^{\gamma}_{~\alpha \beta}(Q)+K^{\gamma}_{~\alpha \beta}(T) \;,
		\label{decomposition_connection}
	\end{equation}
	where $K^\gamma_{~\alpha \beta}(T)$ is the contorsion tensor that depends solely on torsion, whilst $J^\gamma_{~\alpha \beta}(Q)$ is the disformation tensor that is a function of  non-metricity only. The requirement $\nabla_\alpha g_{\mu \nu}\vert_{J^{\gamma}_{~\alpha \beta}=0}=0$ leads to
	\begin{equation}
		K_{\alpha \beta \gamma}=\frac{1}{2}(T_{\alpha \beta \gamma}+T_{\beta \alpha \gamma}+T_{\gamma \alpha \beta}) \;,
		\label{contorsion_torsion}
	\end{equation}
	and imposing  $\Gamma^\gamma_{~[\alpha \beta]}\vert_{K_{\alpha \beta \gamma}=0}=0$ implies
	\begin{equation}
		J_{\alpha \mu \nu}=\frac{1}{2}(Q_{\alpha \mu \nu}-Q_{\nu \alpha \mu}-Q_{\mu \alpha \nu}) \;.
		\label{disformation}
	\end{equation}
Since torsion and non-metricity each carry three tensor indices, it is convenient to split them further into vector- and pure tensor-parts. For torsion, we obtain \cite{Hehl:1994ue,Obukhov:1997zd,Shapiro:2001rz}:
	\begin{align}
		&\text{the trace vector:\ } T^{\alpha}  =g_{\mu \nu}T^{\mu \alpha \nu} \;, \label{torsionTrace}\\
		&	\text{the pseudo trace axial vector:\ } \hat{T}^{\alpha}=\epsilon^{\alpha \beta \mu \nu}T_{\beta \mu \nu} \;, \label{torsionAxial}\\
		&\text{the pure tensor part:\ } t^{\alpha \beta \gamma}  \text{\ that satisfies\ } g_{\mu \nu}t^{\mu \alpha \nu}=0=\epsilon^{\alpha \beta \mu \nu}t_{\beta \mu \nu} \label{torsionTensor} \;.
	\end{align}
	Torsion can be be reconstructed in terms of these irreducible pieces as:
	\begin{equation}
		T_{\alpha \beta \gamma }= -\frac{2}{3}g_{\alpha [\beta}T_{\gamma]}+\frac{1}{6}\epsilon_{\alpha \beta \gamma \nu}\hat{T}^\nu +t_{\alpha \beta \gamma} \;.
		\label{irrep_torsion}
	\end{equation}
	Similarly, we can split further non-metricity into three contributions \cite{Hehl:1994ue,Obukhov:1997zd}:\footnote
	{We remark that this decomposition is not irreducible since $q^{\alpha \beta \gamma}$ can be further decomposed into a fully symmetric part and a remainder.}
	\begin{align}
		&	\text{a first vector:\ } Q^\gamma=g_{\alpha \beta }Q^{\gamma \alpha \beta}\;, \label{nonMetricityVector1}\\
		&\text{a second vector:\ } \hat{Q}^\gamma=g_{\alpha \beta}Q^{\alpha \gamma \beta}\;, \label{nonMetricityVector2}\\
		&\text{the pure tensor part:\ } q^{\alpha \beta \gamma} \text{\ that satisfies\ } g_{\alpha \beta }q^{\gamma \alpha \beta}=0=g_{\alpha \beta}q^{\alpha \gamma \beta} \;. \label{nonMetricityTensor}
	\end{align}
	In terms of the components of \eqs \eqref{nonMetricityVector1} to \eqref{nonMetricityTensor}, non-metricity can be decomposed as follows:
	\begin{equation}
		Q_{\alpha \beta \gamma}= \frac{1}{18}[g_{\beta \gamma}(5Q_{\alpha}-2\hat{Q}_{\alpha})+2g_{\alpha(\beta}(4\hat{Q}_{\gamma)}-Q_{\gamma)})]+ q_{\alpha \beta \gamma} \;.
		\label{irrep_metricity}
	\end{equation}
	Correspondingly, we can express contorsion and disformation as: 
	\begin{align}
		K_{\alpha \beta \gamma} & = \frac{1}{12}\epsilon_{\alpha \beta \gamma \delta}\hat{T}^\delta -\frac{2}{3} g_{\beta [\alpha}T_{\gamma]}+2 t_{[\alpha|\beta|\gamma]}\;, \label{contorsionIrrep}\\
		J_{\alpha \beta \gamma}&=\frac{1}{9}g_{\alpha (\beta}\hat{Q}_{\gamma)}+\frac{1}{18}g_{\beta \gamma}(-5\hat{Q}_\alpha+\frac{7}{2}Q_\alpha) -\frac{5}{18}g_{\alpha (\beta}Q_{\gamma)}+2 q_{\alpha (\beta \gamma)} \;.\label{disformationIrrep}
	\end{align}
Those expressions, like all equations involving decomposition into irreducible components, can be verified using the companion Mathematica notebook \cite{Rigouzzo_Zell_2023}. 

Moreover, we can split curvature \eqref{RicciScalar} as \cite{Rigouzzo:2022yan}:
	\begin{equation}
		\begin{split}
			R&=\mathring{R}+\mathring{\nabla}_\alpha (Q^\alpha - \hat{Q}^\alpha +2 T^\alpha) -\frac{2}{3}T_\alpha (T^\alpha+Q^\alpha-\hat{Q}^\alpha)  +\frac{1}{24}\hat{T}^\alpha \hat{T}_\alpha + \frac{1}{2} t^{\alpha \beta \gamma} t_{\alpha \beta \gamma } \\
			&-\frac{11}{72} Q_\alpha Q^\alpha+ \frac{1}{18}\hat{Q}_\alpha \hat{Q}^\alpha+ \frac{2}{9} Q_\alpha \hat{Q}^\alpha+\frac{1}{4}q_{\alpha \beta \gamma}(q^{\alpha \beta \gamma}-2 q^{\gamma \alpha \beta}) +t_{\alpha \beta \gamma } q^{\beta \alpha  \gamma} \;.
		\end{split}
		\label{curvatureSplit}
	\end{equation}
We shall briefly point out that $R$ is invariant under projective transformations \cite{Schroedinger:1950,Trautman1973,Sandberg:1975db,Hehl:1976kv,Trautmann1976,Hehl:1978, Hehl:1981}
\begin{equation}
	\Gamma^\gamma_{~\alpha \beta} \rightarrow \Gamma^\gamma_{~\alpha \beta} + \delta^\gamma_\beta A_\alpha  \;,
	\label{projective}
\end{equation}
where $A_{\alpha}=A_{\alpha}(x)$ is an arbitrary covariant vector field. The geometric vectors transform as (see \cite{Rigouzzo:2021,Rigouzzo:2022yan}):
\begin{equation} \label{projective_irreps}
	T^\alpha \rightarrow  T^\alpha + 3 A^\alpha, \quad \hat{T}^\alpha \rightarrow \hat{T}^\alpha, \quad
	Q^\alpha \rightarrow Q^\alpha -8 A^\alpha, \quad	\hat{Q}^\alpha \rightarrow \hat{Q}^\alpha -2 A^\alpha \;.
\end{equation}
Thus, generic terms composed of torsion and non-metricity are not projectively invariant -- only their specific combination contained in $R$ is.

	So far, we have considered the basis induced by the metric $g_{\mu\nu}$, to which we refer by Greek indices and that transforms covariantly under diffeomorphisms. Alternatively, one can use a second (internal) basis, that we label with capital Latin indices and that exhibits covariance under local Lorentz transformation, whilst the metric is fixed to being Minkowski $\eta_{AB}$ \cite{Carroll:2004}. We can go from one basis to the other via tetrads $e^\mu_A$, \ie $V^\mu=e^\mu_A V^A$ and inversely $V^A=e^A_\mu V^\mu$ for a vector $V^A$. In particular, we have
	\begin{equation}
		\eta_{AB} = e^\mu_A e^\nu_B g_{\mu\nu} \qquad \Leftrightarrow \qquad g_{\mu\nu} = e_\mu^A e_\nu^B \eta_{AB} \;.
	\end{equation}
	The form of the covariant derivative is different for the two bases. While the affine connection $\Gamma^\alpha_{\mu \nu}$ acts on spacetime indices, the spin connection $w_\mu^{AB}$ corresponds to the internal basis:
	\begin{equation} \label{covariantDerivatives}
		\nabla_\mu V^\nu= \partial_\mu V^\nu + \Gamma^\nu~_{\mu \alpha}V^\alpha  \; , \qquad  \qquad \nabla_\mu V^A= \partial_\mu V^A + \omega_{\mu \; \: B}^{~A} V^B \;.
	\end{equation}
	Compatibility of the covariant derivative in the two bases, \ie $\nabla_\mu V^\nu = e^\nu_A \nabla_\mu V^A$, implies\footnote
	{Alternatively, it is also possible to consider two independent connections \cite{Raatikainen:2019qey}.}
	\begin{equation}
		\omega_{\mu \; \: B}^{~A}=e_\nu^{~A} e^\lambda_{~B} \Gamma^\nu~_{\mu \lambda}-e^\lambda_{~B}\partial_\mu e_\lambda^{~A} \;.
		\label{spin_connection}
	\end{equation}
Correspondingly, the projective transformation \eqref{projective} acts as
	\begin{equation} \label{projective_spin}
			\omega_{\mu \; \: B}^{~A} \rightarrow 	\omega_{\mu \; \: B}^{~A} + \delta^A_{~B} A_\mu \;,
\end{equation}
\ie only a part symmetric in the Latin indices is generated.
\\According to \eq \eqref{decomposition_connection}, we can split the spin connection in its Riemannian part,
	\begin{equation} \label{spin_connection_riemann}
		\mathring{\omega}_{\mu \; \: B}^{~A}=e_\nu^{~A} e^\lambda_{~B} \mathring{\Gamma}^\nu_{~\mu \lambda}-e^\lambda_{~B}\partial_\mu e_\lambda^{~A} \;,
	\end{equation}
	and contributions due to torsion and non-metricity:
	\begin{equation} \label{spin_connection_split}
		\omega_{\mu \; \: B}^{~A} = \mathring{\omega}_{\mu \; \: B}^{~A} + e_\nu^{~A} e^\lambda_{~B} J^\nu_{~\mu \lambda} + e_\nu^{~A} e^\lambda_{~B} K^\nu_{~\mu \lambda} \;.
	\end{equation}
		The symmetry of contorsion, $K_{\alpha\beta\gamma}=K_{[\alpha|\beta|\gamma ]}$, implies that:\footnote{The implication from the right to the left can be verified by noticing that imposing $J_{(\alpha|\beta|\gamma)}=0$ implies $J_{(\alpha|\beta|\gamma)} = -\frac{1}{2}Q_{\beta \alpha \gamma}=0$.}
	\begin{equation} \label{antisymmetric_spin_connection}
		Q_{\gamma \alpha \beta} = 0 \qquad \Longleftrightarrow \qquad \omega_{\mu A B} = \omega_{\mu [A B]} \;.
	\end{equation}
	We conclude that the spin connection is antisymmetric in the Latin indices if and only if non-metricity vanishes.

	\subsection{Coupling to fermions and vector fields}
	
	An important reason for introducing the spin connection $w_{\mu}^{~AB}$ is that it is necessary to couple fermions to gravity. As detailed in appendix \ref{app:localLorentz} (see \eg also \cite{Diakonov:2011fs,Magueijo:2012ug,Donoghue:2016vck}), the covariant derivative of a spinor field $\Psi$ reads\footnote
	{Alternatively, one can also attempt to construct a fermionic covariant derivative without anti-symmetrization in the gamma matrices; in this case, projective invariance is lost and even a minimal coupling to gravity leads to an interaction with non-metricity \cite{Demir:2022wpp}.}
	\begin{equation} \label{covariant_derivative_fermion}
		\mathcal{D}_\mu \Psi= \partial_\mu \Psi +\frac{1}{8} \omega_\mu^{~AB}\left(\gamma_A\gamma_B - \gamma_B\gamma_A\right) \Psi \;.
	\end{equation}
	Now we can plug in the split \eqref{spin_connection_split} of the spin connection. We use $\gamma^\mu \gamma^\nu \gamma^\rho=-g^{\mu \nu} \gamma^\rho-g^{\nu \rho} \gamma^\mu+g^{\mu \rho} \gamma^\nu+i \epsilon^{\sigma \mu \nu \rho} \gamma_\sigma \gamma^5$ to derive from \eqs \eqref{contorsionIrrep} and \eqref{disformationIrrep} that 
	\begin{align}
		K_{\alpha \mu \beta}\gamma^\mu\left(\gamma^\alpha \gamma^\beta- \gamma^\beta \gamma^\alpha \right)&= 2 K_{\alpha \mu \beta } (-2 g^{\alpha \mu} \gamma^\beta + i \epsilon^{\rho  \mu \alpha\beta }\gamma_{\rho}\gamma^{5})= 4 \gamma^\alpha T_\alpha -i \gamma^\alpha \gamma^5 \hat{T}_\alpha \;,\\
		J_{\alpha \mu \beta}\gamma^\mu\left(\gamma^\alpha \gamma^\beta- \gamma^\beta \gamma^\alpha \right) & =J_{\alpha \mu \beta}(-g^{\mu \alpha}\gamma^\beta+g^{\mu \beta}\gamma^\alpha)=  2 \gamma^\alpha(Q_\alpha -\hat{Q}_\alpha) \;.
	\end{align}
Next we consider a generic kinetic term \cite{Freidel:2005sn,Alexandrov:2008iy} of the fermionic field and decompose it into its Riemann part and contributions due to torsion and non-metricity:
	\begin{equation} \label{develop_fermion}
		\begin{split}
			&\frac{i}{2}\bar{\Psi}(1-i\alpha-i\beta \gamma^5)\gamma^\mu \mathcal{D}_\mu \Psi + \text{h.c.}\\
			= & \frac{i}{2}\bar{\Psi}\gamma^\mu \mathcal{\mathring{D}}_\mu \Psi  + \frac{i}{16}\bar{\Psi}(1-i\alpha-i\beta \gamma^5)(4\gamma^\alpha T_\alpha-i \gamma^\alpha \gamma^5 \hat{T}_\alpha+2\gamma^\alpha(Q_\alpha-\hat{Q}_\alpha))\Psi + \text{h.c.}\\
				=&\left(\frac{i}{2}\bar{\Psi}\gamma^\mu \mathcal{\mathring{D}}_\mu \Psi + \text{h.c.}\right) -\frac{1}{8} \bar{\Psi} \gamma^5  \gamma^\alpha  \Psi \hat{T}_\alpha\\
				& +\frac{1}{4}\alpha\left(2 T_\alpha + Q_\alpha-\hat{Q}_\alpha\right) \bar{\Psi}\gamma^\alpha  \Psi  +\frac{1}{4}\beta\left(2 T_\alpha + Q_\alpha-\hat{Q}_\alpha\right) \bar{\Psi} \gamma^5 \gamma^\alpha   \Psi    \;.
		\end{split}
	\end{equation}
First, we can discuss a minimal coupling, $\alpha=\beta=0$. In this case, we reproduce the well-known result that fermions only couple to the axial part $\hat{T}_\alpha$ of torsion \cite{Kibble:1961ba} and that no interaction with non-metricity arises (see \cite{Adak:2008gd, BeltranJimenez:2017tkd, BeltranJimenez:2018vdo, Koivisto:2018aip, Jimenez:2019woj, Delhom:2020hkb, BeltranJimenez:2020sih}).\footnote
{Further discussions about the coupling of fermions to the torsionful teleparallel formulation of GR \cite{Einstein:1928, Einstein:19282, Moller:1961, Pellegrini:1963, Hayashi:1967se,Cho:1975dh, Hayashi:1979qx} can be found in  \cite{Hayashi:1979qx, deAndrade:1997cj, deAndrade:1997gka, deAndrade:2000kr, deAndrade:2001vx, Obukhov:2002tm,Maluf:2003fs, Mielke:2004gg,Mosna:2003rx, Obukhov:2004hv, Aldrovandi:2013wha}.}
In contrast, we observe that  non-zero $\alpha$ and $\beta$ generically lead to a coupling of fermions to both torsion vectors $T_\alpha$ and $\hat{T}_\alpha$ as well as to both non-metricity vectors $Q_\alpha$ and $\hat{Q}_\alpha$. 

Finally, we shall comment on projective invariance of the fermionic kinetic term. Since a projective transformation only generate a part symmetric in the internal indices (see \eq \eqref{projective_spin}), it is clear that the covariant derivative \eqref{covariant_derivative_fermion} is unaffected even in the presence of non-minimal couplings. This agrees with the result \eqref{develop_fermion}, where projective invariance can be verified by plugging in the transformations \eqref{projective_irreps}. Because of this projective symmetry, non-minimal couplings of fermions as shown in \eq \eqref{develop_fermion} cannot distinguish between effects of the parity-even torsion vector $T_\alpha$ and a certain combination of the non-metricity vectors $Q_\alpha$ and $\hat{Q}_\alpha$. This can lead to a certain universality with regard to the different formulation of GR, as exemplified in section \ref{ssec:metricAffinePortal}.

At this point, we can discuss two approaches for coupling fermions two gravity.
\paragraph*{Option 1: ``Gravity first''.} The first option is to start from gravity, as we have also done in our presentation. In this case, one must begin by making a choice about the geometry of gravity, \ie select which of the three features curvature $R$, torsion $T^\alpha_{\ \beta \gamma}$  and non-metricity $Q_{\alpha\beta \gamma}$ are present. Subsequently, one can decide how gravity couples to fermions. As is evident from \eq \eqref{develop_fermion} (see also appendix \ref{sapp:covderivfermion}), only the interaction with the Riemannian spin connection $\mathring{\omega}_{\mu}$ is necessary for ensuring gauge invariance under local Lorentz transformations. In contrast, there is a freedom concerning the coupling with torsion and non-metricity. 
\paragraph*{Option 2: ``Matter first''.} An alternative point of view is to start from matter fields in flat spacetime. In this case, one can derive GR by gauging the Poincar\'e group \cite{Utiyama:1956sy,Kibble:1961ba,Sciama:1962}, which is significantly more constraining. First, it leads to the EC formulation, \ie curvature and torsion arise whereas non-metricity does not appear, as evident from appendix \ref{sapp:covderivfermion}. Second, gauging the Poincar\'e group can only generate a kinetic term of the form \eqref{develop_fermion} and in the simplest case only a minimal coupling arises, where $\alpha=\beta=0$.

In the following, we shall mainly follow the first approach, although our result encompass as special case also the predictions of the second option. It is important to emphasize that although \textit{non-metricity cannot be regarded as a consequence of gauging the Poincar\'e group, its coupling to fermions is fully consistent.}

\paragraph*{Vector fields.}	Finally, we make a brief remark about the use of covariant derivatives when coupling vector fields to gravity. First, we consider an Abelian gauge field $\tilde{A}_\mu$. In Riemannian geometry, partial and covariant derivatives are interchangeable since
	\begin{equation} 
		\mathring{D}_{[\mu}  \tilde{A}_{\nu]} = \partial_{[\mu} \tilde{A}_{\nu]} \;,
	\end{equation}
	where we used that $\mathring{\Gamma}^\alpha_{~[\mu \nu]} \tilde{A}_\alpha =0$. However, this changes once the Christoffel symbol is no longer symmetric in its lower indices. In this case, using a covariant derivative is inconsistent since a coupling  $T^\alpha_{\mu\nu} \tilde{A}_\alpha$ would break invariance under gauge transformations $\tilde{A}_{\nu} \rightarrow \tilde{A}_{\nu} + \partial_\nu \epsilon$. Therefore, the field strength tensor of an Abelian gauge field in metric-affine gravity reads (see \eg \cite{Hehl:1976kj, Shapiro:2001rz, Karananas:2021zkl}):
	\begin{equation} \label{fieldStrength}
		F_{\mu\nu} = 2 \partial_{[\mu} \tilde{A}_{\nu]} \;.
	\end{equation}
	Analogously, also non-Abelian gauge fields must not couple directly to $\Gamma^\alpha_{~\mu \nu}$.

	\subsection{Selection rules}
	\label{ssec:selectionrules}
	As detailed in the introduction, our goal is to consider gravitational theories that are as general as possible while still being equivalent to metric GR in the absence of matter. In order to achieve this, criteria for constructing an action of gravity coupled to matter were already devised in \cite{Karananas:2021zkl}. We shall develop further the conditions of \cite{Karananas:2021zkl}, where we employ the derivative counting put forward in \cite{Diakonov:2011fs}. We propose that the action should obey the following two conditions:
	\begin{enumerate}
		\item[I.] There should be no operator with more than two derivatives, where the connection $\Gamma_{~\alpha \beta}^\gamma$ counts as one derivative.
		\item[II.] The mass dimension of any operator should not be greater than $4$. 
	\end{enumerate}
	Here we assume that the action is analytic and view it as polynomial in gravitational and matter fields. Additionally, we require Lorentz invariance and locality as well as invariance under all gauge groups in the matter sector. We shall briefly outline our motivation for choosing the two criteria listed above -- a more detailed discussion can be found in appendix \ref{app:criteria}.
	
	We begin by commenting on the derivative counting contained in condition I.). That the Levi-Civita connection $\mathring{\Gamma}_{~\alpha \beta}^\gamma$ features one derivative is evident from \eq \eqref{levicivita}. As discussed in \cite{Diakonov:2011fs}, this represents a first reason for also counting the full connection $\Gamma_{~\alpha \beta}^\gamma$ as one derivative. Then the decomposition \eqref{decomposition_connection} of the connection implies that also contorsion $K^{\gamma}_{~\alpha \beta}$ as well as disformation $J^{\gamma}_{~\alpha \beta}$ correspond to one derivative. In turn, this is equivalent to associating one derivative to torsion $T_{\gamma\alpha\beta}$ and to non-metricity $Q_{\gamma\alpha\beta}$.
	
	The purpose of restricting ourselves to two derivatives in condition I.) is to ensure that the coupling of gravity and matter does not lead to new propagating degrees of freedom in addition to those already present in the matter sector and the two polarizations of the massless graviton. Including more than two derivatives generically causes the presence of additional propagating degrees of freedom, although there are exceptions. A particular consequence of criterion II.) is that the matter sector is renormalizable. This implies that non-renormalizable effects only arise due to gravity, \ie curvature, torsion and non-metricity. Since GR is not renormalizable independently of the chosen formulation, this can be viewed as a minimal approach to non-renormalizability.

	\section{Standard Model coupled to metric-affine gravity}
	\label{sec:Gen-Action}
		\subsection{The theory}
	Having set up our selection rules, we can now write down the most general Lagrangian. In the gravity sector we consider the generic metric-affine formulation of GR, where curvature, torsion and non-metricity are a priori present. Then we wish to couple it to all types of matter present in the SM: scalar, fermion and vector fields. For now the only restriction that we impose is selection rule I), limiting the number of derivatives.
	Then the most general action following the criteria listed above can be decomposed in three parts:
	\begin{equation}\label{general_action_component}
		S= S_{\text{metric}} +S_{\text{sources}}+S_{\text{quadratic}} \;,
	\end{equation}
	with 
	\begin{equation} \label{action_metric}
		\begin{split}
			S_{\text{metric}}= &\int \mathrm{d}^{4} x \sqrt{-g} \Big[\frac{1}{2} \Omega^2 \mathring{R}-V(\Phi,\Psi, F_{\mu \nu})\\
			&  - \tilde{K}_1 g^{\alpha \beta} \mathring{D}_{\alpha} \Phi \mathring{D}_{\beta}  \Phi^\star+\frac{i\tilde{K}_2}{2} ( \bar{\Psi}\gamma^\mu \mathring{D}_\mu \Psi- \overline{\mathring{D}_\mu \Psi} \gamma^\mu \Psi)-\frac{1}{4}\tilde{K}_3g^{\mu \alpha} g^{\nu \beta}F_{\mu \nu} F_{\alpha \beta} \Big] \;,
		\end{split}
	\end{equation}
	\begin{equation} \label{action_sources}
		S_{\text{sources}}= \int \mathrm{d}^{4} x \sqrt{-g}\Big[-J_1^\alpha \hat{T}_\alpha - J_2^\alpha T_\alpha - J_3^\alpha \hat{Q}_\alpha - J_4^\alpha  Q_\alpha \Big] \;,
	\end{equation}
	and 
	\begin{equation} \label{action_quadratic}
		\begin{split}
			S_{\text{quadratic}}= &\int \mathrm{d}^{4} x \sqrt{-g}\Big[B_{1} Q_{\alpha} Q^{\alpha}+B_{2} \hat{Q}_{\alpha} \hat{Q}^{\alpha}+B_{3} Q_{\alpha} \hat{Q}^{\alpha}+B_{4} q_{\alpha \beta \gamma} q^{\alpha \beta \gamma}+B_{5}q_{\alpha \beta \gamma} q^{\beta \alpha \gamma}\\
			&+C_{1}T_{\alpha} T^{\alpha}+C_{2} \hat{T}_{\alpha} \hat{T}^{\alpha}+C_{3} T_{\alpha} \hat{T}^{\alpha} +C_{4} t_{\alpha \beta \gamma}t^{\alpha \beta \gamma}\\
			&+D_1\epsilon_{\alpha \beta \gamma \delta}t^{\alpha \beta \lambda}t^{\gamma \delta}_{~~\lambda}+D_2\epsilon_{\alpha \beta \gamma \delta}q^{\alpha \beta \lambda}q^{\gamma \delta}_{~~\lambda}+D_{3}\epsilon_{\alpha \beta \gamma \delta}q^{\alpha \beta \lambda}t^{\gamma \delta}_{~~\lambda}\\
			&+E_1 T_{\alpha}Q^\alpha+E_2\hat{T}_{\alpha}Q^\alpha+E_3 T_{\alpha}\hat{Q}^\alpha+E_4\hat{T}_\alpha \hat{Q}^\alpha+E_5 t^{\alpha \beta \gamma}q_{\beta \alpha \gamma} \Big] \;.
		\end{split}
	\end{equation}
	Here $S_{\text{metric}}$ gathers all term that are torsion- and non-metricity-free and therefore would also exist in the metric formulation of GR. All sources for torsion and non-metricity are in $S_{\text{sources}}$, and terms that are quadratic in torsion and/or non-metricity are collected in $S_{\text{quadratic}}$. Finally, let us emphasize that the theory considered here is purely classical.
	
Note that in the second line of \eq \eqref{action_metric}, $\mathring{D}_{\alpha}$ denotes the Riemannian covariant derivative, which contains couplings to gauge fields. In the case of fermions, $\mathring{D}_\mu \Psi$, as defined in \eq \eqref{covariant_derivative_fermion}, moreover includes the Riemannian spin connection  $\mathring{\omega}_{\mu ~  B}^{~ A}$ (see \eq \eqref{spin_connection_riemann}). In contrast, the field strength tensor $F_{\alpha \beta}$, which is defined in \eq \eqref{fieldStrength}, only contains partial derivatives. To simplify notations we did not indicate the dependency of the functions in front of the term, except for the potential $V(\Phi, \psi, F_{\mu \nu})$. At this point, without applying selection rule II, all of these functions may depend on the complex scalar field $\Phi$ and on the fermion $\psi$, but not on the field strength $F_{\mu \nu}$. This is due to the fact that we consider an action with no more than two derivatives, following selection rule I).

\subsection{Equivalent metric theory}
	As the equations of motions are algebraic, solutions can be explicitly found. Since there are no source terms for the pure tensor parts $t_{\alpha \beta \gamma}$ and $q_{\alpha \beta \gamma}$ in \eq \eqref{action_sources} (see discussion in appendix \ref{sapp:inclu_matter}), they have to vanish:
	\begin{equation}
		t_{\alpha \beta \gamma}=q_{\alpha \beta \gamma}=0 \;.
	\end{equation}
 Once we find the solutions for $T^\alpha$, $\hat{T}^\alpha$, $Q^\alpha$ and $\hat{Q}^\alpha$, we can plug them back into the action  \eqref{action_sources} and \eqref{action_quadratic} to obtain an equivalent action. The details of these computations, such as solutions of equation of motion and conformal transformation of the metric and fermionic field, can be found in appendix \ref{app_details_computations}. Picking up all pieces, the final Lagrangian reads: 
\begin{equation} \label{final_lagrangian}
	\begin{split}
		S & = \int \mathrm{d}^{4} x \sqrt{-g} \Big[\frac{1}{2}  \mathring{R}-\frac{V(\Phi,\Psi, F_{\mu \nu})}{\Omega^4} - \frac{3}{\Omega^2}g^{\alpha \beta} \mathring{D}_{\alpha}\Omega \mathring{D}_{\beta}\Omega- \frac{\tilde{K}_1}{\Omega^2} g^{\alpha \beta} \mathring{D}_{\alpha} \Phi \mathring{D}_{\beta}  \Phi^\star \\
		&+\frac{i\tilde{K}_2}{2 \Omega^3} ( \bar{\Psi}\gamma^\mu \mathring{D}_\mu \Psi- \overline{\mathring{D}_\mu \Psi} \gamma^\mu \Psi)-\frac{1}{4}\tilde{K}_3g^{\mu \alpha} g^{\nu \beta}F_{\mu \nu} F_{\alpha \beta} +\frac{1}{\Omega^2}\sum_{i\leq j=1}^{4}\mathcal{L}_{ij} J^\alpha_{i} J_{j,\alpha}\Big] \;.
	\end{split}
\end{equation}
In this form, we have integrated out torsion and non-metricity, the effects of which are completely mapped to the sum of ten source-squared terms, weighted by the $\mathcal{L}_{ij}$ coefficients shown in appendix \ref{sapp_solving} (\eq \eqref{L_ij}). Via a conformal transformation of the metric, we also went to the Einstein frame where gravity is minimally coupled to matter. This resulted in a modification of the potential term, of the scalar and fermionic kinetic terms and the apparition of new geometry induced interaction terms.

Although the discussion of our paper is classical, we remark that integrating out torsion $T_{\alpha \beta \gamma}$ and non-metricity $Q_{\alpha \beta \gamma}$ is also possible on the quantum level. The reason is that our initial action \eqref{general_action_component} only contains terms that are at most quadratic in $T_{\alpha \beta \gamma}$ and $Q_{\alpha \beta \gamma}$. Therefore, the path integrals with respect to $T_{\alpha \beta \gamma}$ and $Q_{\alpha \beta \gamma}$ are Gaussian and can be carried out exactly by plugging the result of the classical equations of motion back into the action (see \eg \cite{Weinberg:1995mt} and discussion in \cite{Shaposhnikov:2020geh}). The only caveat is that the quantum theory can feature additional contributions arising from the measure of the path integral (which cannot even be defined unless a measure is specified for the initial action \eqref{general_action_component} -- see \cite{Aros:2003bi} for details about the path integral measure in different formulations of gravity).

We are now in a good position to impose the selection rule II. By limiting the mass dimension of interactions between matter and gravity to $4$, this criterion restricts heavily the form the couplings can take:\footnote{Evidently, the restriction to mass dimension $4$ only applies in the initial Jordan frame, not the Einstein frame where matter is minimally coupled to gravity. This is the essence of our approach, in which only specific higher dimension operators are generated due to the deviation from Riemannian geometry.}
\begin{equation}
	\begin{split}
		&\tilde{K}_i=k_{i0}  \;, \qquad  \Omega^2(\varphi)=f_0+\xi \varphi^2 \;, \qquad D_i(\varphi)=d_{i0}+d_{i1} \varphi^2\,, \qquad i=1,2,3 \;, 
		\\& 
		J_j^\alpha = a_j^{(\varphi)} \mathring{D}^\alpha \varphi^2 + a_j^{(\Phi)} S^\alpha +  a_j^{V} V^\alpha  +  a_j^{A}A^\alpha, \qquad
		C_j(\varphi)=c_{j0}+c_{j1}\varphi^2 \;,  \qquad j=1,2,3,4 \;,\\
		& B_k(\varphi)=b_{k0}+b_{k1}\varphi^2 \;, \qquad E_k(\varphi)=e_{k0}+e_{k1}\varphi^2 \;, \qquad  k=1,2,3,4,5 \;.
	\end{split}
	\label{selec_rulE_3}
\end{equation}
where we also took into account the discrete symmetry $\varphi\rightarrow - \varphi$ and defined the currents:\footnote
{Cf.\@\xspace \eqs \eqref{complexScalarSource}, \eqref{vectorFermionSource} and \eqref{axialFermionSource} in appendix \ref{sapp:inclu_matter} (see also \eq  \eqref{sourceGeneral}).}
\begin{align}\label{new_currents_1}
	S_\alpha &  =\frac{i}{2}\left(\Phi^\star \mathring{\mathcal{D}}_\alpha \Phi - (\mathring{\mathcal{D}}_\alpha \Phi)^\star \Phi \right) \;,\\ \label{new_currents_2}
	V_\alpha &= J^{(V)}_\alpha= \bar{\Psi} \gamma_\alpha \Psi \;,\\ \label{new_currents_3}
	A_\alpha &=  J^{(A)}_\alpha	=\bar{\Psi} \gamma^5 \gamma_\alpha \Psi \;,
\end{align}
Moreover, $\varphi$ corresponds to the real part of the complex scalar field $\Phi$:
\begin{equation}
	\varphi=\sqrt{2}|\Phi| \; .
\end{equation} 
\\Let us discuss briefly the form of the terms above:
\begin{itemize}
	\item The kinetic terms are already of mass dimension $4$, hence $\tilde{K}_i(h)$ can only be a dimensionless constant. Note that we can set the constant $k_{i0}=1$ without loss of generality by a redefinition of the scalar, fermionic and vector field.
	\item The scalar curvature $R$ can only be coupled to $\varphi^2$ because $R$ carries mass dimension $2$, and the only scalar of mass dimension $2$ one can construct is $\varphi^2$. For example, terms involving two spinor fields are discarded because they have mass dimension $3$. One can also set $f_0=1$ by a general field redefinition.
	\item All the quadratic couplings $B$'s, $C$'s, $D$'s and $E$'s can accommodate both a constant and a $\varphi^2$-term.
	\item The $4$-vector currents are comprised of four possibles terms (see also \eq \eqref{sourceGeneral} in appendix \ref{sapp:inclu_matter}).
	\item In terms of the parameters used in \eq \eqref{develop_fermion}, we would obtain \cite{Karananas:2021zkl}\footnote
	{The sign of $a_1^{A}$ differs as compared to \cite{Karananas:2021zkl}. (This can be absorbed by altering the convention \eqref{epsilonConvention} of the Levi-Civita tensor.) As a result, some terms in the equivalent metric theory derived in \cite{Shaposhnikov:2020frq} change sign.\label{fnSignDifference}}
	\begin{equation} \label{alpha_beta_mapping}
		\begin{split}	a_1^{V} = 0 \;, \qquad a_1^{A} = \frac{1}{8} \;,  \qquad 	a_2^{V} = -\frac{\alpha}{2}  \;, \qquad a_2^{A} = -\frac{\beta}{2} \;,\\
			a^V_3=\frac{\alpha}{4}	 \;, \qquad a^A_3=\frac{\beta}{4} \;, \qquad a^V_4=-\frac{\alpha}{4}  \;, \qquad a^A_4=-\frac{\beta}{4}
			\;.
		\end{split}
	\end{equation}
\end{itemize}

	Let us now substitute all the other expressions in the action. By plugging in \eq \eqref{selec_rulE_3} into the expressions for the $\mathcal{L}_{ij}$ coefficients given in \eq \eqref{L_ij} and performing an additional conformal transformation of the fermion field, we obtain the full equivalent theory (see \eqs \eqref{eq1} to \eqref{eq10} in appendix \ref{sapp:interaction_terms}): 
	\begin{align} \label{eqAA}
		\mathcal{L}_{AA} &= \Omega^2 \frac{\sum_{n=0}^3 P^{(AA)}_n \varphi^{2 n}}{D} A_{\mu}A^\mu \;,\\ \label{eqVV}
		\mathcal{L}_{VV} &= \Omega^2 \frac{\sum_{n=0}^3 P^{(VV)}_n \varphi^{2 n}}{D}V_{\mu}V^\mu \;,\\ 
		\label{eqhh}
		\mathcal{L}_{\varphi \varphi} &=\Omega^{-2} \frac{\sum_{n=0}^3 P^{(\varphi \varphi)}_n \varphi^{2 n}}{D} \mathring{D}^\mu \varphi^2\mathring{D}_\mu \varphi^2;,\\
		\mathcal{L}_{\Phi \Phi}&=\Omega^{-2} \frac{\sum_{n=0}^3 P^{(\Phi \Phi)}_n \varphi^{2 n}}{D}S^\mu S_\mu  ;,\\ \label{eqAV}
		\mathcal{L}_{AV} &= 2\Omega^{2}\frac{\sum_{n=0}^3 P^{(AV)}_n \varphi^{2 n}}{D}A_{\mu}V^\mu \;,\\ \label{eqAh}
		\mathcal{L}_{A \varphi} &=   2\frac{\sum_{n=0}^3 P^{(A\varphi)}_n \varphi^{2 n}}{D}A_{\mu}\mathring{D}^\mu \varphi^2 \;,\\
		\mathcal{L}_{A \Phi} &= 2 \frac{\sum_{n=0}^3 P^{(A \Phi)}_n \varphi^{2 n}}{D}A_{\mu}S^\mu \;,\\ \label{eqVh}
		\mathcal{L}_{V \varphi} &= 2 \frac{\sum_{n=0}^3 P^{(V \varphi)}_n \varphi^{2 n}}{D}V_{\mu}\mathring{D}^\mu \varphi^2 \;,\\
		\mathcal{L}_{V \Phi} &=  2\frac{\sum_{n=0}^3 P^{(V \Phi)}_n \varphi^{2 n}}{D}V_{\mu}S^\mu \;,\\
		\mathcal{L}_{\varphi \Phi} &= 2\Omega^{-2} \frac{\sum_{n=0}^3 P^{( \varphi \Phi)}_n \varphi^{2 n}}{D}S_{\mu}\mathring{D}^\mu \varphi^2 \;, \label{eqSh}
	\end{align}
	with 

		\begin{equation}
			D= \sum_{m=0}^4 O_m \varphi^{2 m} \;.
	\end{equation}
	All terms have the same form, namely a fraction of polynomials of $\varphi$. The numerator is a polynomial of order $6$ whilst the denominator is the same for all expressions and is of order $8$. The coefficients $P_n$ are identical up to the subscript $(AA)$, $(VV)$ etc. The difference simply amounts to the $a_i$'s coefficients displayed in \eq \eqref{eq1} to \eq \eqref{eq10}. So overall, there are $5$ independent polynomials in the common denominator and $4$ in each numerator. The exact form of the polynomials can be found using the companion Mathematica notebook \cite{Rigouzzo_Zell_2023}.
	Then the final action can be expressed as: 	
		\begin{equation} \label{final_final_lagrangian}
			\begin{split}
				S & = \int \mathrm{d}^{4} x \sqrt{-g} \Big[\frac{1}{2}  \mathring{R}-\frac{V(\Phi,\Omega^{3/2}\Psi, F_{\mu \nu})}{(1+\xi \varphi^2)} - \frac{3\xi^2}{4(1+\xi \varphi^2)^2}g^{\alpha \beta} \mathring{D}_{\alpha}\varphi^2 \mathring{D}_{\beta}\varphi^2- \frac{1}{1+\xi \varphi^2} g^{\alpha \beta} \mathring{D}_{\alpha} \Phi \mathring{D}_{\beta}  \Phi^\star \\
				&+\frac{i}{2} ( \bar{\Psi}\gamma^\mu \mathring{D}_\mu \Psi- \overline{\mathring{D}_\mu \Psi} \gamma^\mu \Psi)-\frac{1}{4}g^{\mu \alpha} g^{\nu \beta}F_{\mu \nu} F_{\alpha \beta} + \mathcal{L}_{AA}+\mathcal{L}_{VV}+\mathcal{L}_{\varphi \varphi}+\mathcal{L}_{\Phi \Phi} \\
				& +\mathcal{L}_{AV}+\mathcal{L}_{A\varphi}+\mathcal{L}_{V\varphi}+\mathcal{L}_{V\Phi}+\mathcal{L}_{\varphi \Phi}	\Big] \;.
			\end{split}
		\end{equation}
	It encompasses the results of \cite{Karananas:2021zkl} and \cite{Rigouzzo:2022yan} as special cases, as we discuss in appendix \ref{sapp:knownlimits}.

Let us now analyze the numbers of parameters in the action for each field of the Standard Model in the presence of torsion and non-metricity. Looking at each part of the action, we can make the following comments:
	\begin{itemize}
		\item The quadratic action $S_\text{quadratic}$ defined in \eq \eqref{action_quadratic} only couples to the real scalar field. Indeed, after imposing selection rule II), all the functions $A_i,B_i,C_i,D_i,E_i$ can only depend on the scalar field $\Phi$, see \eq \eqref{selec_rulE_3}. Since there are two parameters in each of the function and \eq \eqref{action_quadratic} contains $17$ quadratic terms, $34$ parameters arise.
		\item The action $S_\text{sources}$ defined in \eq \eqref{action_sources} features sources for the complex scalar field as well as for the fermion. As can be seen in the definition \eqref{new_currents_1} of currents, there are $2$ currents for the complex scalar field -- one for the real part and one for the complex part -- and $2$ currents for the fermion -- one for the vector part and one for the axial part. Each of these currents exist in four copies, corresponding to the four irreducible vector representations of torsion and non-metricity. 
		\item Finally, we can discuss the metric action $S_\text{metric}$ defined in \eq \eqref{action_metric}. In this part of the action, we do not consider the kinetic term as an independent coupling because we can always canonicalize them by a field redefinition. By the same token, we do not consider the Planck mass in this parameter counting, because it is dimensionful whereas all other couplings to torsion and non-metricity can be chosen to be dimensionless.
		Therefore, the only term that results in an independent parameter is the non-minimal coupling of the scalar curvature to the scalar field. 
	\end{itemize} 
		Therefore, we see that each fields does not contribute equally to the parameter counting in the presence of torsion and non-metricity. For example, a complex scalar field comes with $17+1+8=26$ independent parameter whilst a fermionic field comes with $8$ coupling to torsion and non-metricity. A gauge vector field does not couple independently to torsion or non-metricity, given our selection rules. The contributions are summarized in tables \ref{summary_indep_terms} and \ref{summary_indep_terms_without_pure_irrep}, where we also included the parameter counting for the Standard Model. We assumed for simplicity that fermions couple universally to gravity (see also \cite{Shaposhnikov:2020aen}). Since there is one complex scalar field, the Higgs boson, adding all contribution together gives $34+16+1=51$. Since there is always the possibility of a global rescaling of the fractions shown in \eqs \eqref{eqAA} to \eqref{eqSh}, we have to subtract one coupling constant so the total number of independent parameters is $50$.
		\begin{table}[H]
			\centering
			\begin{tabular}{|l|c|c|c|c|}
				\hline
				& $S_\text{quadratic}$ & $S_\text{sources}$ & $S_\text{metric}$ & Total \\ \hline
				Pure Gravity  &   17                   & 0                   & 0                &   17\\ \hline
				Complex scalar field        &   17                   & 8                   & 1                   & 26      \\ \hline
				Fermionic field             &   0                   &8                    &        0           &8       \\ \hline
				Gauge vector field          &0                      &          0          & 0                  &0       \\ \hline
				 Standard Model & $17 +17=34$ &  $8+ 8=16$                 &  $1$                  &                   $34+16+1=51$      \\ \hline
			\end{tabular}
		\caption{Summary of the number of independent couplings in metric affine theory for different particle content.}
		\label{summary_indep_terms}
		\end{table}

\begin{table}[H]
	\centering
	\begin{tabular}{|l|c|c|c|c|}
		\hline
		& $S_\text{quadratic}$ & $S_\text{sources}$ & $S_\text{metric}$ & Total \\ \hline
		Pure Gravity  &   10                 & 0                   & 0          &   10 \\ \hline
		Complex scalar field        &   10                 & 8                   & 1                   & 19      \\ \hline
		Fermionic field             &   0                   &8                    &        0           &8       \\ \hline
		Gauge vector field          &0                      &          0          & 0                  &0       \\ \hline
	Standard Model & $10+10=20$ &  $8+ 8=16$                 &  $1$                  &                   $20+16+1=37$      \\ \hline
	\end{tabular} 
	\caption{Summary of the number of independent couplings in metric affine theory for different particle content, after taking into account that $t_{\alpha \beta \gamma}=q_{\alpha \beta \gamma}=0$.}
	\label{summary_indep_terms_without_pure_irrep}
\end{table}

	\section{Implications for production of fermionic dark matter}
	\label{sec:darkmatter}
	\subsection{Review of Einstein-Cartan portal to dark matter}
	As reviewed in the introduction, the choice of formulation of GR can have important implications for dark matter. We shall discuss this in the scenario of \cite{Shaposhnikov:2020aen}, where a fermion $\Psi_N$, which is a singlet under all gauge groups of the SM, is added to the field contents of the SM. The observation of neutrino masses provides a strong motivation for considering such a scenario since a mixing of singlet fermions with active neutrinos can generate masses for the latter via the well-known seesaw mechanism \cite{Minkowski:1977sc,Yanagida:1979as,Gell-Mann:1979vob,Mohapatra:1979ia,Schechter:1980gr,Schechter:1981cv}  (see \cite{Boyarsky:2018tvu} for a review). For the present discussion, however, it is inessential if $\Psi_N$ couples directly to SM-neutrinos, provided the mixing is small enough to obey observational constraints, \eg due to bounds on X-rays signals from radiatively decaying dark matter \cite{Boyarsky:2018tvu}. 
	In the metric formulation of GR, production mechanisms arising from a mixing of $\Psi_N$ with active neutrinos have been proposed \cite{Dodelson:1993je,Shi:1998km}. Here the difficulty consists in producing a sufficient abundance of dark matter while at the same time satisfying the upper bounds on the strength of mixing coming from X-ray constraints. Whereas this is not possible for the so-called non-resonant production \cite{Dodelson:1993je}, a resonant production \cite{Shi:1998km} can be implemented  \cite{Shaposhnikov:2008pf}. However, this requires a fine-tuning of parameters \cite{Canetti:2012vf,Canetti:2012kh,Ghiglieri:2020ulj} and constrains the mass of $\Psi_N$ is a narrow range \cite{Boyarsky:2018tvu}. 
	
	In \cite{Shaposhnikov:2020aen}, a gravitational generation of $\Psi_N$ was studied in the EC formulation of GR. This mechanism was dubbed \textit{Einstein-Cartan portal to dark matter} and its starting point are non-minimal fermionic kinetic terms for both $\Psi_N$ and  the fermions of the SM (see \eq \eqref{develop_fermion}):
	\begin{equation} \label{nonMinimalKineticTerms}
		\frac{i}{2}\bar{\Psi}(1-i\alpha-i\beta \gamma^5)\gamma^\mu \mathcal{D}_\mu \Psi + \text{h.c} \;,
	\end{equation}
	where we assume that the coupling to gravity is universal, \ie identical for all fermions.
	After solving for torsion, the resulting equivalent metric theory features new four-fermion interactions \cite{Shaposhnikov:2020frq} (\cf \eqs \eqref{eqAA}, \eqref{eqVV} and \eqref{eqAV})
	\begin{align}
		\mathcal{L}_{AA}& \simeq \frac{3\beta^2-3}{16} A_\mu A^\mu \;,\label{AAEC} \\ 
		\mathcal{L}_{VV} &\simeq   \frac{3\alpha^2}{16}V_{\mu}V^\mu \;, \label{VVEC}\\ 
		\mathcal{L}_{AV} &\simeq  \frac{3\alpha \beta}{8}A_{\mu}V^\mu \;. \label{AVEC}
	\end{align} 
	Since the current $V_\mu$ and $A_\mu$ contain sums over all fermion species, processes arise in which two fermions of the SM annihilate and produce two $\Psi_N$-particles. This leads to the abundance $\Omega_N$ of $\Psi_N$ \cite{Shaposhnikov:2020aen}:
	\begin{equation}
		\frac{\Omega_N}{\Omega_{DM}}\simeq 3.6 \cdot 10^{-2} \, C_f \left( \frac{M_N}{10~\text{keV}}\right) \, \left( \frac{T_\mathrm{prod}}{M_P} \right)^3 \;,
		\label{general_abundance}
	\end{equation}
	where $\Omega_{DM}$ is the observed abundance of dark matter, $M_N$ is the mass of $\Psi_N$ and $T_\mathrm{prod}$ represents the highest temperature at which $\Psi_N$ are produced, such as the reheating temperature of a hot Big Bang. Moreover, the factor $C_f$ is sensitive to whether $\Psi_N$ is Dirac or Majorana:
	\begin{align}
		C_D & = \frac{9}{4}\bigg\{45\left( 1+\alpha^2-\beta^2 \right)^2 + 21 \left( 1- \left(\alpha + \beta\right)^2 \right)^2 + 24 \left( 1- \left(\alpha - \beta\right)^2 \right)^2\bigg\} \;,\\
		C_M &= \frac{9}{4}\bigg\{24\left( 1+\alpha^2-\beta^2 \right)^2 + 21 \left( 1- \left(\alpha + \beta\right)^2 \right)^2\bigg\} \;,
	\end{align}
	corresponding to the former and latter case, respectively. It is evident from \eq \eqref{general_abundance} that depending on the values of the non-minimal couplings $\alpha$ and $\beta$, the observed abundance of dark matter can be generated in a wide range of masses, from $M_N\sim 1\, \text{keV}$ to $M_N\sim 10^8\, \text{GeV}$, and without a fine-tuning of parameters. 
	\\Four comments are in order regarding this Einstein-Cartan portal to dark matter.
	\begin{itemize}
		\item The equivalent metric theory considered in \cite{Shaposhnikov:2020aen} also contains scalar-fermion interactions of the form shown in \eqs \eqref{eqAh} and \eqref{eqVh}. This leads to a second channel for producing $\Psi_N$, namely from the annihilation of two Higgs particles. For certain parameter choices, this process can lead to a higher abundance of $\Psi_N$ than the one due to four-fermion interaction displayed in \eq \eqref{general_abundance} \cite{Shaposhnikov:2020aen}. In the following, we shall focus on the generation of $\Psi_N$ from fermion annihilation.
		\item If a connection to neutrinos is made, the mixing of $\Psi_N$ with active neutrinos has to be sufficiently small due to X-ray constraints discussed above as well as the requirement that dark matter should be approximately stable on cosmological timescales \cite{Boyarsky:2006jm, Shaposhnikov:2020aen}. As a result, the contribution of $M_N$ to the mass splitting of active neutrinos is negligible and two other sterile neutrinos are required to generate the observed mass differences. This can be achieved in the framework of the $\nu$MSM, which extends the particle content of the SM by three right-handed neutrinos \cite{Asaka:2005an,Asaka:2005pn} (see \cite{Boyarsky:2009ix} for a review). One of them, which can be identified with $\Psi_N$, acts a dark matter whereas the mass splittings arise from the other two right-handed neutrinos. Moreover, the latter can generate the observed baryon asymmetry in our Universe via leptogenesis \cite{Kuzmin:1985mm,Fukugita:1986hr} (see \cite{Boyarsky:2009ix,Drewes:2013gca,Drewes:2017zyw,Klaric:2021cpi,Abdullahi:2022jlv} for reviews and recent developments). Because of the small mixing of $\Psi_N$, a prediction of the $\nu$MSM is that the lightest active neutrino is practically massless \cite{Asaka:2005an,Asaka:2005pn}. This statement still holds in the whole range of $M_N$ that arises from the Einstein-Cartan portal to dark matter \cite{Shaposhnikov:2020aen}.
		\item The lower bound on $M_N$ around $1\, \text{keV}$ arises since hot dark matter is incompatible with observed small scales in the power spectrum, as \eg inferred from Lyman-$\alpha$ measurements (see \cite{Boyarsky:2018tvu}). In contrast, a value of $M_N$ around a few keV, corresponding to warm dark matter, has interesting observational consequences. First, the produced $\Psi_N$-particles have a characteristic momentum distribution, which may serve to confirm or exclude this proposal of producing dark matter due to the effects of gravity \cite{Shaposhnikov:2020aen}. Secondly, the non-negligible momenta that arise in this scenario may fit Lyman-$\alpha$ data better than pure cold dark matter \cite{Garzilli:2018jqh,Garzilli:2019qki}.
		\item The Einstein-Cartan portal to dark matter is independent of a possible phase of inflation since only a sufficiently high temperature $T_\mathrm{prod}$ is required in the early Universe. However, it is interesting to consider this gravitational production of fermions after a period of inflation driven by the Higgs field \cite{Bezrukov:2007ep}. This proposal of Higgs inflation is highly sensitive to the choice of formulation of GR \cite{Bauer:2008zj,Rasanen:2018ihz, Raatikainen:2019qey, Langvik:2020nrs, Shaposhnikov:2020gts}. We shall consider the Palatini version \cite{Bauer:2008zj} since it has favorable properties with regard to quantum corrections \cite{Bauer:2010jg} (see discussions in \cite{Shaposhnikov:2020fdv,Karananas:2022byw,Poisson:2023tja}). In this scenario, choosing universal energy scales in the scalar and fermionic sector leads to $\alpha\sim\beta\sim \sqrt{\xi}$, where $\xi \approx 10^7$ \cite{Shaposhnikov:2020aen}. Then the observed abundance of dark matter is produced for $M_N$ around a few keV, which leads to the observationally interesting case of warm dark matter discussed above. 
	\end{itemize}
	
	\subsection{Portal to dark matter in metric-affine and Weyl gravity}
	\label{ssec:metricAffinePortal}
	Now we shall study a gravitational production of fermions in the metric-affine formulation of GR. We consider the case in which the scalar field $\varphi$ has a negligible expectation value. This is true even if we identify $\varphi$ with the Higgs field and take into account scenarios of Higgs inflation, as long as inflation has ended and reheating has been completed. When we only keep the leading terms in small $\varphi$, the new interaction terms induced by the effect of classical torsion and non-metricity (see \eqs \eqref{eqAA}, \eqref{eqVV}, \eqref{eqAV} \eqref{eqAh} and \eqref{eqVh}), reduce to:
	\begin{align}
		\mathcal{L}_{AA} &\simeq \frac{P_0^{(AA)}}{O_0} A_\mu A^\mu \;, \quad \mathcal{L}_{VV} \simeq \frac{P_0^{(VV)}}{O_0} V_\mu V^\mu \;, \quad \mathcal{L}_{AV} \simeq 2 \frac{P_0^{(AV)}}{O_0} A_\mu V^\mu \;, \label{FourFermionMAG}\\
		\mathcal{L}_{A \varphi} &\simeq 2 \frac{P_0^{(A \varphi)}}{O_0} A_\mu D^\mu \varphi^2 \;, \quad \mathcal{L}_{V \varphi} \simeq 2 \frac{P_0^{(A\varphi)}}{O_0} A_\mu D^\mu \varphi^2 \;. \label{FourFermionMAG2}
	\end{align}
All expressions for the polynomials in the numerator and denominator can be found using \cite{Rigouzzo_Zell_2023} and are also displayed in appendix \ref{subsec:portaltoDM}.

	First, we shall consider the limit of EC gravity. 
	In the absence of non-metricity, the previous interaction terms reduce to: 
	\begin{align} \label{limitAA}
		\mathcal{L}_{AA}& \simeq \frac{c_{10} (a_1^{A})^2+c_{20} (a_2^{A})^2-c_{30} a_1^A a_2^A}{c_{30}^2-4 c_{10} c_{20}} A_\mu A^\mu \;, \\ \label{limitVV}
		\mathcal{L}_{VV} &\simeq   \frac{c_{10} (a_1^{V})^2+c_{20} (a_2^{V})^2-c_{30} a_1^V a_2^V}{c_{30}^2-4 c_{10} c_{20}}V_{\mu}V^\mu \;,\\ \label{limitAV}
		\mathcal{L}_{AV} &\simeq \frac{2c_{10} a_1^Aa_1^V+2c_{20} a_2^{A}a_2^V -c_{30} (a_2^A a_1^V+a_1^A a_2^V)}{ c_{30}^2-4 c_{10} c_{20}}A_{\mu}V^\mu \;.
	\end{align}
	If we specialize momentarily to the non-minimal fermionic kinetic terms shown in \eq \eqref{develop_fermion}, then \eq \eqref{alpha_beta_mapping} fixes $a_1^V$, $a_1^A$, $a_2^V$ and $a_2^A$ in terms of $\alpha$ and $\beta$. Moreover, a purely gravitational action that only consists of the term $\frac{1}{2} R$ leads to $c_{10}=-\frac{1}{3}$, $c_{20}=\frac{1}{48}$ and $c_{30}=0$, as is evident from \eq \eqref{curvatureSplit}. Plugging these choices back into \eqs \eqref{limitAA} to \eq \eqref{limitAV}, we reproduce the result of \cite{Shaposhnikov:2020frq} as displayed in \eqs \eqref{AAEC} to \eqref{AVEC}.  
	
	Now we shall go beyond this specific choice of parameters. First, already in EC gravity the coupling of fermions to gravity generically features more than the two parameters $\alpha$ and $\beta$. Namely, there are four a priori independent coupling constants even in the absence of non-metricity (see \eq \eqref{selec_rulE_3}). This makes the three coefficients of the $A_\mu A^\mu$-, $V_{\mu}V^\mu$- and $A_{\mu}V^\mu$-interactions, which are shown in \eqs \eqref{limitAA} to \eqref{limitAV}, independent. While a quantitative analysis of such a situation remains to be performed, it is clear that the qualitative features of the Einstein-Cartan portal to dark matter remain unchanged even for three independent four-fermion interactions.
	
	Next, we can discuss other formulations of GR. As noted in \cite{Rigouzzo:2022yan}, EC gravity, which features curvature and torsion, and Weyl gravity, in which curvature and non-metricity are present, are equivalent for analogous choices of source terms. As a result, the gravitational production of fermionic dark matter can equally well be implemented in the Weyl formulation.
	Only a small difference arises when instead of coupling fermionic currents to irreducible representations of torsion and non-metricity, one employs a non-minimal kinetic term \eqref{nonMinimalKineticTerms} since in this case the coupling of $\hat{T}$ to fermions has no counterpart in Weyl gravity (see \eq \eqref{develop_fermion}). Instead of \eqs \eqref{AAEC} to \eqref{AVEC}, one gets 
		\begin{align}\label{limitAAWeyl}
	\mathcal{L}_{AA}& \simeq \frac{b_{10} (a_3^{A})^2+b_{20} (a_4^{A})^2-b_{30} a_3^A a_4^A}{b_{30}^2-4 b_{10} b_{20}} A_\mu A^\mu = \frac{3 \beta^2}{16}A_\mu A^\mu \;, \\ \label{limitVVWeyl}
	\mathcal{L}_{VV} & \simeq \frac{b_{10} (a_3^{V})^2+b_{20} (a_4^{V})^2-b_{30} a_3^V a_4^V}{b_{30}^2-4 b_{10} b_{20}}V_{\mu}V^\mu=\frac{3\alpha^2}{16}V_\mu V^\mu \;,\\ \label{limitAVWeyl}
	\mathcal{L}_{AV} & \simeq \frac{2b_{10} a_3^Aa_4^V+2b_{20} a_3^{A}a_4^V -b_{30} (a_4^A a_3^V+a_3^A a_4^V)}{ b_{30}^2-4 b_{10} b_{20}}A_{\mu}V^\mu= \frac{3 \alpha \beta}{8}A_\mu V^\mu \;.
	\end{align}
For sufficiently large $\alpha$ and $\beta$, these operators have the same effect as their counterparts \eqref{AAEC}, \eqref{VVEC} and \eqref{AVEC} in EC gravity. Once independent couplings of fermions to torsion and/or non-metricity are included, corresponding to \eq \eqref{selec_rulE_3}, EC and Weyl gravity -- and hence their respective portals to fermionic dark matter -- are fully equivalent.
	
	Finally, we go to metric-affine gravity, which encompasses both the EC and Weyl formulations as special cases. For the same choice of action as in \cite{Shaposhnikov:2020aen}, we can exploit the fact that $R$ and the covariant derivative of fermions are projectively invariant to set the combination $Q_\alpha - \hat{Q}_\alpha$ to zero. This removes the coupling of fermions to non-metricity (see \eq \eqref{develop_fermion}) and so we recover the result of the EC formulation as shown in \eqs \eqref{AAEC} to \eqref{AVEC}. Thus, these interactions can equivalently be viewed as predictions of metric-affine gravity, where no assumption is required about the vanishing of non-metricity. If instead we allow for more general choices of coupling constants, the four-fermion interactions in EC gravity, as displayed in \eqs \eqref{limitAA} to \eqref{limitAV}, will also be indistinguishable from those in metric-affine gravity shown in \eqref{FourFermionMAG}:
 Dark matter is produced from the three possible four-fermion interactions, mediated by $A_\mu A^\mu$-, $V_{\mu}V^\mu$- and $A_{\mu}V^\mu$-terms, which come with three a priori independent coefficients. Nevertheless, some quantitative features may be different as compared to the situation investigated in \cite{Shaposhnikov:2020aen}, both due to the independence of the four-fermion interactions and the importance of the scalar-fermion channel mentioned above. We leave a detailed investigation of these questions for the future.
	
	We have demonstrated that the gravitational portal to dark matter proposed in \cite{Shaposhnikov:2020aen} is not specific to the EC formulation of GR. It also exists in Weyl and metric-affine gravity with qualitatively the same features: A production of fermionic dark matter can be implemented in a wide range of masses and without a finetuning of parameters. Moreover, a characteristic primordial momentum distribution arises which can be used to probe this mechanism in the case of warm dark matter. \textit{In summary, one can regard a gravitational portal to fermionic dark matter as a generic feature of GR.} In other words, the absence of this mechanism in metric gravity appears as a peculiarity specific to this formulation of GR.

\section{Conclusion}
\label{sec:conclusion}	
Our current understanding of fundamental physics rests on two pillars, namely the Standard Model (SM) of particle physics and General Relativity (GR) as the description of gravity. Far-reaching puzzles arise when the two are combined. Already at the classical level, a question has remained unanswered about which formulation of GR one should employ. Since all of them are fully equivalent in pure gravity, they have to be regarded as equivalent incarnations of one and the same theory and thus represent an inherent ambiguity of GR. Once GR couples to matter such as in the SM, the various versions of GR are no longer equivalent and generically lead to distinct observables predictions. Thus, it is important to explore how concrete effects depend on the formulation of GR.

To this end, we have employed metric-affine gravity. This formulation of GR is special for two reasons. First, no assumptions are required about the geometry of gravity. Instead, curvature, torsion and non-metricity  are determined dynamically through the principle of stationary action. Second, this version of GR encompasses the metric, Palatini, EC and Weyl formulations as special cases and so the results obtained in metric-affine gravity can be carried over automatically to these other incarnations of GR.

In a first step, we discussed the coupling of gravity to different matter fields. (Complex) scalar fields and fermions generically interact with  both torsion and non-metricity, sourcing these two geometric features. In contrast, there is no independent coupling for vector fields because of their gauge symmetry. Subsequently, we refined the criteria of \cite{Karananas:2021zkl} for constructing an action of matter interacting with GR. The goal of such selection rules is to avoid assumptions as far as possible while still ensuring equivalence to the metric formulation of GR in the absence of matter. In particular, they preserve the property that the only propagating degree of freedom is the massless spin 2 graviton. Correspondingly, our criteria that we developed in  section \ref{ssec:selectionrules} aim at capturing the minimal ambiguity that is inevitably contained in GR. 

Using these selection rules, we subsequently constructed in \eqs \eqref{action_metric} to \eqref{action_quadratic} a model for coupling (complex) scalar fields, fermions and gauge bosons to metric-affine gravity. In this way, we extended our study \cite{Rigouzzo:2022yan}, in which we only considered a real scalar field, to encompass all fields of the SM. Subsequently, we brought our theory to an equivalent form in the metric formulation of GR, which leads to new geometry-induced interaction terms stemming from the underlying presence of torsion and non-metricity. Together with the corresponding Mathematica code \cite{Rigouzzo_Zell_2023}, this lays the groundwork for investigating phenomenological implications.

As a first example, we discussed a new mechanism for producing fermionic dark matter candidates, which was first discovered in the EC formulation \cite{Shaposhnikov:2020aen}. We showed that this gravitational portal to dark matter is not unique to EC gravity but also exists in the Weyl and metric-affine versions of GR with very similar predictions. This universality partly arises as consequence of projective invariance of a non-minimal coupling of fermions to gravity. Therefore, observational signatures of this gravitational dark matter production, such as a characteristic primordial momentum distribution, can serve to distinguish metric gravity from other formulations of GR. 
Evidently, further approaches are needed to constrain and ultimately determine the formulation of GR that is realized in Nature. 

Arguably, a precise understanding of classical GR and the ambiguity resulting from its equivalent formulations represents a necessary prerequisite for tackling some of the open issues of quantum gravity. Indeed, parts of the parameter space of our model can lead to an inflationary generation of primordial inhomogeneities \cite{Bezrukov:2007ep,Bauer:2008zj,Rasanen:2018ihz, Langvik:2020nrs,Shaposhnikov:2020gts,Rigouzzo:2022yan} (see also \cite{Raatikainen:2019qey}). Moreover, our theory is capable of addressing -- via a non-perturbative gravitational effect -- the long-standing question of why the Higgs vacuum expectation value is so much smaller than the Planck scale \cite{Shaposhnikov:2020geh}. Our findings may have implications for further unresolved problems related to quantum gravity, such as infrared phenomena \cite{Weinberg:1965nx,Carney:2017jut,Gomez:2018war},
a possible running of Newton's constant \cite{Weinberg:1980,Berges:2000ew,Reuter:2004nv} and the potential breakdown of the classical metric description \cite{Dvali:2013eja, Dvali:2017eba,Michel:2023ydf}.
	\acknowledgments
	
It is a pleasure to thank Will Barker, Georgios Karananas, Roberto Percacci and Misha Shaposhnikov for useful comments and discussions.
C.R.~acknowledges support from the Science and Technology Facilities Council (STFC). S.Z.~acknowledges support of the Fonds de la Recherche Scientifique -- FNRS.

	\appendix
		
	\section{Details about local Lorentz symmetry}
	\label{app:localLorentz}
	\subsection{Covariant derivative of fermions}
	\label{sapp:covderivfermion}
	We shall review the construction of the covariant derivative of a fermion in gravity. Our starting point is the spin connection $\omega_\mu^{\ AB}$ as introduced in \eq \eqref{covariantDerivatives} via the derivative of a vector $V^A$: 
	\begin{equation}
		\nabla_\mu V^A= \partial_\mu V^A + \omega_{\mu \; \: B}^{~A} V^B \;.
	\end{equation}
	In the non-coordinate basis, one is free to perform a Lorentz transformation as it preserves the form of the metric: $\eta_{A'B'}=\Lambda_{A'}^A \Lambda_{B'}^B \eta_{AB}$. Importantly, the Lorentz transformation can be different at every point, making it local.
	Since a local Lorentz transformation $\Lambda^{A'}_{\ A}(x)$ acts as $V^{A'} = \Lambda^{A'}_{\ A}(x) V^A$, requiring covariance of $\nabla_\mu V^A$ implies \cite{Carroll:2004}
	\begin{equation} \label{transformationSpinConnection}
		\omega_{\mu \ B'}^{\ A'} = \Lambda^{A'}_{\ A} \Lambda^B_{\ B'} 	\omega_{\mu \ B}^{\ A} - \Lambda^{C}_{\ B'} \partial_\mu \Lambda^{A'}_{\ C} \;,
	\end{equation}
	where we drop the $x$ dependence in the Lorentz transformation in the following.
	Crucially, we observe that the inhomogeneous part of the transformation is anti-symmetric in the internal indices. This holds true even if $\omega_\mu^{\ (AB)}\neq 0$, \ie if non-metricity is present. Moreover, we note that $\omega_\mu^{\ AB}$ transforms homogeneously under diffeomorphisms. 
	Now we consider an infinitesimal transformation,
	\begin{equation}
		\Lambda^{A}_{\ B} = \delta^{A}_{\ B} + \omega^{A}_{\ B} \;,
	\end{equation}
	where $\omega_{BA} = - \omega_{AB}$ are the generators of the Lorentz group. Then \eq \eqref{transformationSpinConnection} yields
	\begin{equation} \label{transformationSpinConnectionInfinitesimal}
		\delta \omega_{\mu A B} = \omega_A^{\ C} \omega_{\mu C B} + \omega_B^{\ C} \omega_{\mu AC} - \partial_\mu \omega_{AB} \;.
	\end{equation}
	A generic element $\Lambda$ of the Lorentz group can be represented as \cite{Peskin:1995ev}: 
	\begin{equation}
		\Lambda=e^{-\frac{i}{2} \omega_{AB} J^{AB}} \;, \label{Poincare}
	\end{equation}
	and the generators $J^{AB}_R$ determine the transformation of a representation $\phi^i$:
	\begin{equation}
		\delta \phi^i=-\frac{i}{2} \omega_{A B}\left(J_R^{AB}\right)^i{ }_j \phi^j \;.
	\end{equation}
	For example, for a four-vector we have \cite{Peskin:1995ev}:
	\begin{equation}
		\left(J^{\mu \nu}_\text{vector}\right)_\sigma^\rho=i\left(\eta^{\mu \rho} \delta_\sigma^\nu-\eta^{\nu \rho} \delta_\sigma^\mu\right) \;,
	\end{equation} 
	whereas a Dirac spinor corresponds to:
	\begin{equation}
		\left(J^{A B}_\text{spinor}\right)_\sigma^\rho=\frac{i}{4} (\gamma^A \gamma^B-\gamma^B \gamma^A)_\sigma^\rho \;,
		\label{representation_spinor}
	\end{equation} 
	with the gamma matrices $\gamma^A$ as defined in \eq \eqref{gammaConvention}. Note that the factor of $\frac{i}{4}$ is the right convention to ensure that the generators satisfy the Lorentz algebra: 
	\begin{equation}
		\left[J^{\kappa \lambda}, J^{\mu \nu}\right]=-i g^{\lambda \mu} J^{\kappa \nu}+i g^{\kappa \nu} J^{\mu \lambda}+i g^{\lambda \nu} J^{\kappa \mu}-i g^{\kappa \mu} J^{\nu \lambda} \;.
		\label{lorentzalgebra}
	\end{equation}
	For a spacetime-dependent $\omega_{AB}(x)$, the partial derivative of a Dirac spinor therefore transforms as
	\begin{equation} \label{partialFermion}
		\partial_\mu \psi^{\prime}=	\partial_\mu \left(\Lambda \psi(x) \right) = \Lambda \left( \partial_\mu \psi(x) + \frac{1}{8} \partial_\mu \omega_{AB} (\gamma^A\gamma^B-\gamma^B\gamma^A) \psi(x)\right) \;.
	\end{equation}
	Therefore, covariance is lost for $\partial_\mu \omega_{AB}\neq 0$. We can use the spin connection $\omega_\mu^{\ AB}$ to restore covariance by defining
	\begin{equation}
		D_\mu \psi=\partial_\mu \psi +\frac{1}{8} \omega_\mu^{~AB}(\gamma_A \gamma_B-\gamma_B \gamma_A) \psi \;.
	\end{equation}
	Indeed, it follows from \eq \eqref{transformationSpinConnectionInfinitesimal} that an infinitesimal local Lorentz transformation yields 
	\begin{equation}
		\delta\left(D_\mu \psi-\partial_\mu \psi\right)^{\prime}  = - \frac{1}{8} \partial_\mu \omega_{AB} (\gamma^A \gamma^B-\gamma^B \gamma^A) \psi (x) \;,
	\end{equation}
	which precisely cancels the inhomogeneous contribution of \eq \eqref{partialFermion}.
	
	\subsection{Field strength in Einstein-Cartan gravity}
	\label{sapp:fieldStrength}
	In the argument presented above, we have used the form \eqref{covariantDerivatives} as an input and then demonstrated covariance. As is well-known, one can also proceed in the other direction and derive gravity from gauging the Poincar\'e group \cite{Utiyama:1956sy,Kibble:1961ba,Sciama:1962}. In this case, the tetrad and spin connection emerge as gauge fields associated to translations and the Lorentz group, respectively. As a result, one obtains  the EC formulation of GR, which only features curvature and torsion emerge so that the spin connection is anti-symmetric, $\omega_{\mu}^{~AB} = \omega_{\mu}^{~[AB]}$. Focusing on Lorentz transformations, we can draw a direct parallel with other gauge theories: When a symmetry becomes local, derivative terms are generically no longer covariant. One needs to introduce a gauge field, the variation of which is the derivative of the parameters of the gauge group. In table \ref{gauge_theory}, we show an overview of different Abelian and non-Abelian gauge groups to highlight their similarities.
	\begin{table}
		\begin{center}
			\begin{tabular}{|c|c|c|c|}
				\hline
				Local Lie group     & Local parameter & Gauge field & Variation of Gauge field \\ \hline \hline
				U(1)     & $e$          &    $A_\mu$         &       $\delta A_\mu= \partial_\mu e$                       \\ \hline
				SU(3)   & $g^i \qquad i=1,...,8$        &    $A^i_\mu$         &       $\delta A^i_\mu= \partial_\mu g^i $                        \\ \hline
				Lorentz &            $\omega_{AB}$     &    $\omega_{\mu}^{~AB}$         &            $\delta \omega_{\mu}^{~AB}=\partial_{\mu}\omega^{AB}$          \\ \hline
			\end{tabular}
			\caption{Comparison of different gauge groups and their corresponding gauge fields and field strengths.}
		\label{gauge_theory}
		\end{center}
	\end{table}

	Evidently, the gauge field itself is not gauge invariant, only the field strength is. For Abelian gauge theories, like $U(1)$, the field strength is $F_{\mu \nu}=2\partial_{[\mu}A_{\nu]}$ whilst for non-Abelian gauge theories like $SU(3)$ we have $G^i_{\mu \nu}=2\partial_{[\mu}A^i_{\nu]}+ f^{iab}A^a_\mu A^b_\nu$, where $f^{iab}$ are the structure constants of $SU(3)$ \cite{Rubakov:2002fi}. Likewise, for the Lorentz group, the field strength (which corresponds to the Riemann tensor), is defined as 
	\begin{equation} \label{gaugeFieldStrength}
		R^{AB}_{\mu \nu}=2\partial_{[\mu}\omega^{AB}_{\nu]} +\frac{i}{2} f^{ABCDEF}\omega^{CD}_{[\mu }\omega^{EF}_{\nu] } \;.
	\end{equation}
The factor of $\frac{i}{2}$ comes from the convention chosen for the element of the Lorentz group in \eqref{Poincare}, and $f^{ABCDEF}$ are the structure constants of the Lorentz group. They can be read off from the commutation relation of the Lorentz generators \cite{Peskin:1995ev}:
	\begin{equation}
		[J^{AB},J^{CD}]=f^{ABCDEF}J_{EF} \;,
	\end{equation}
	and are given by: 
	\begin{equation}
		f^{ABCDEF}=-2i(\eta^{[B|C}\eta^{A]E}\eta^{FD}-\eta^{[B|D}\eta^{A]E}\eta^{CF}) \;.
		\label{structureconstant}
	\end{equation}
	Plugging \eqref{structureconstant} in \eqref{gaugeFieldStrength}, we evaluate the formula for the Riemann tensor (see also \cite{Karananas:2021zkl}):
	\begin{equation}
		R^{AB}_{\mu \nu}=2\partial_{[\mu}\omega^{AB}_{\nu]}+2\omega_{[\mu}^{~BF}w^{~A}_{\nu]~F} \;,
	\end{equation}
where we used the antisymmetry of $\omega_{\mu}^{~AB}$. For EC gravity, we can now reproduce \eq \eqref{riemannTensor}, namely the Riemann tensor with four Greek indices using the relationship \eqref{spin_connection} between the spin connection and affine connection:
\begin{equation} \label{riemannSpin}
	R^{\alpha \beta}_{~\mu \nu}=e^\alpha_A e^\beta_B R^{AB}_{~\mu \nu}=e^\alpha_A e^\beta_B\left( 2\partial_{[\mu}\omega^{AB}_{\nu]}+2\omega_{[\mu}^{~BF}w^{~A}_{\nu]~F}\right) \\ 
	= 2 \partial_{[\mu}\Gamma^{\alpha~\beta}_{~\nu] }+2 \Gamma^{\alpha}_{~[\mu| \lambda} \Gamma^{\lambda~\beta}_{~\nu]}  \;.
\end{equation}

Finally, let us comment on the symmetry properties of the Riemann tensor. In the EC formulation, it is evident from \eq \eqref{riemannSpin} that $\omega_{\mu}^{~AB}=\omega_{\mu}^{~[AB]}$ implies that the Riemann tensor is anti-symmetric in the first two indices. We compare this to the situation in other version of GR in table \ref{ll}.  
	\begin{table}[h]
		\begin{center}
			
			\begin{tabular}{|c|c|c|c|}
				\hline
				& Metric & \begin{tabular}[c]{@{}c@{}} Einstein-Cartan\end{tabular} & \begin{tabular}[c]{@{}c@{}} Metric-affine \end{tabular} \\ \hline
				$R_{ab[cd]}$        & \checkmark  & \checkmark                                                                                  & \checkmark                                                                            \\ \hline
				$R_{[ab]cd}$    & \checkmark      & \checkmark                                                                                 & \text{\sffamily X}                                                                           \\ \hline
				$R_{(ab)(cd)}$      & \checkmark      & \text{\sffamily X}                                                                                  & \text{\sffamily X}                                                                    \\ \hline
			\end{tabular}
			\caption{Existence of symmetry properties of Riemann tensor for different formulations of GR.}
			\label{ll}
		\end{center}
	\end{table}

		\section{Discussion of criteria for coupling gravity to matter}
	\label{app:criteria}
	\subsection{Equivalence in pure gravity}
	We shall discuss in more detail the selection rules for coupling gravity to matter as proposed in section \ref{ssec:selectionrules}. First, we consider the situation in the absence of matter. In this case, the action can only feature two kinds of fields. On the one hand, we have the metric $g_{\mu\nu}$, which has mass dimension $0$ and contains no derivative. On the other hand, we have the connection $\Gamma_{~\alpha \beta}^\gamma$ or components thereof, such as torsion and non-metricity. All these fields have mass dimension $1$ and by the derivative counting introduced above, all of them correspond to one derivative. This implies that in pure gravity, the number of derivatives is equal to the mass dimension so that condition I.) implies condition II.). Thus, we only need to consider the former.
	
	Since Riemannian curvature $\mathring{R}$  is quadratic in the connection  $\mathring{\Gamma}_{~\alpha \beta}^\gamma$ (see \eq \eqref{riemannTensor}), criterion I.) amounts to restricting ourselves to terms linear in curvature and quadratic in torsion and non-metricity. Consequently, the most general theory of pure gravity that obeys criterion 1.) is of the form \cite{Rigouzzo:2022yan}
	\begin{equation} \label{actionPureGravitySchematic}
		\mathcal{L} \sim \mathring{R} + c_{TT} T^\mu T_\mu + c_{TQ} T^\mu Q_\mu + c_{QQ} Q^\mu Q_\mu + \ldots \;,
	\end{equation}
	where $c_{TT}$, $c_{TQ}$ and $c_{QQ}$ are arbitrary real parameters and $T_\mu$ and $Q_\mu$ are defined in \eqs \eqref{torsionTrace} and \eqref{nonMetricityVector1}. Only an exemplary selection of possible contributions due to torsion and non-metricity is displayed in \eq \eqref{actionPureGravitySchematic} and the dots represent further terms that are quadratic in torsion and non-metricity and that can be formed using the six tensors $T^\mu$, $\hat{T}^\mu$, $t^{\alpha \beta \gamma}$, $Q^\mu$, $\hat{Q}^\mu$, $q^{\alpha \beta \gamma}$. Each of these contributions comes with an a priori unknown coefficient. Moreover, we left out in \eq \eqref{actionPureGravitySchematic} contributions of the form $\mathring{\nabla}_\mu T^\mu$ since they correspond to full derivatives. 
	
	One can wonder why we included in \eq \eqref{actionPureGravitySchematic} only the Riemann part $\mathring{R}$ of the full curvature $R$. As is evident from \eq \eqref{actionPureGravitySchematic}, the difference of $R$ and $\mathring{R}$ consists of terms that are already present in action. Therefore, replacing $\mathring{R}$ by $R$ can be absorbed by a shift of the coefficients $c_{TT}$, $c_{TQ}$, \etc (see \cite{Karananas:2021zkl} for more details). Conversely, we note that the theory \eqref{actionPureGravitySchematic} has the same structure as the full Ricci scalar $R$, with the only difference that specific coefficients are replaced by arbitrary ones.\footnote
	{More precisely, the model \eqref{actionPureGravitySchematic} contains all contributions quadratic in torsion and non-metricity whereas $R$ only features the subset of terms that are parity-even.}
	Finally, we remark that one can construct another contribution linear in curvature, namely the so-called Holst term \cite{Hojman:19802, Nelson:1980ph, Castellani:1991et, Holst:1995pc}:
	\begin{equation}
		\epsilon^{\mu\nu\rho\sigma}R_{\mu\nu\rho\sigma} =  \frac{1}{3}\hat{T}^\alpha (Q_\alpha-\hat{Q}_\alpha+2 T_\alpha)-\mathring{\nabla}_\alpha \hat{T}^\alpha +\frac{1}{2} \epsilon_{\beta \gamma \delta \lambda}t_\alpha^{~\delta \lambda}t^{\alpha \beta \gamma}+\epsilon_{\alpha \gamma \delta \lambda}q^{\alpha \beta \gamma}t_{\beta}^{~ \delta \lambda}  \;,
	\end{equation}
	which is parity-odd. In analogy to \eq \eqref{curvatureSplit}, we separated contributions due to torsion and non-metricity.
	There is no Riemannian part, $\epsilon^{\mu\nu\rho\sigma} \mathring{R}_{\mu\nu\rho\sigma}=0$, because of the symmetries of $\mathring{R}^\mu_{~\nu\rho\sigma}$. Therefore, including the Holst term in the action would only lead to a shift of some of the coefficients.  
	
	From the Lagrangian \eqref{actionPureGravitySchematic} we can determine torsion and non-metricity by their equations of motions. Since they only appear quadratically, the solution is given by
	\begin{equation}
		T_{\gamma\alpha\beta} = Q_{\gamma\alpha\beta} = 0 \;,
	\end{equation}
	\ie torsion and non-metricity vanish dynamically. Hence, theories of the form \eqref{actionPureGravitySchematic} are equivalent to the metric formulation of GR. This shows that the restriction to two derivatives in criterion I.) ensures equivalence to metric GR in the absence of matter.
	
	Once we include more than two derivatives, the equivalence to GR in the metric formulation is generically lost. If for example we allow for four derivatives, we can among others include terms that are quadratic in the Ricci scalar $\mathring{R}$ or the Ricci tensor $\mathring{R}_{\mu\nu}$. As is well-known, such contributions generically lead to additional propagating degrees of freedom \cite{Stelle:1977ry,Neville:1978bk,Neville:1979rb,Sezgin:1979zf,Hayashi:1979wj, Hayashi:1980qp}.\footnote
	{There are exceptions to this rules -- see \cite{Enckell:2018hmo, Antoniadis:2018ywb, Tenkanen:2019jiq, Gialamas:2019nly, Antoniadis:2019jnz, Lloyd-Stubbs:2020pvx, Antoniadis:2020dfq, Das:2020kff, Gialamas:2020snr, Dimopoulos:2020pas, Karam:2021sno, Lykkas:2021vax, Gialamas:2021enw,  Annala:2021zdt, Dioguardi:2021fmr,Pradisi:2022nmh,Dimopoulos:2022tvn,Dimopoulos:2022rdp,Lahanas:2022mng,Panda:2022can} for studies of corresponding models.}
	Hence, they break the equivalence to metric GR already in the absence of matter. We must mention, however, that one can construct specific theories that do not obey our restriction to two derivatives but nevertheless do not feature any additional propagating degrees of freedom apart from a massless graviton \cite{Karananas:2021zkl}. Thus, criterion I.) is sufficient for ensuring equivalence to the metric formulation of GR but not necessary.
	
	\subsection{Effect of matter}\label{sapp:inclu_matter}
	Next, we discuss the effects of including matter. To begin with, we shall only consider a real scalar field $\varphi$. This situation was already discussed in our previous paper \cite{Rigouzzo:2022yan}, to which we refer the reader for more details. The most general polynomial Lagrangian that obeys criteria I.) and II.) is of the form \cite{Rigouzzo:2022yan}:\footnote
	{The precise form of the action is displayed in \eqs \eqref{general_action_component} till \eqref{action_quadratic}.}
	\begin{equation} \label{actionScalarFieldSchematic}
		\begin{split}
			\mathcal{L} \sim &   \mathring{R} + \xi h^2 \mathring{R} +(c_{TT} + \tilde{c}_{TT} \varphi^2)  T^\mu T_\mu +  (c_{TQ} + \tilde{c}_{TQ} \varphi^2) T^\mu Q_\mu  + (c_{QQ} + \tilde{c}_{QQ}\varphi^2)  Q^\mu Q_\mu\\
			& +\xi_T J_\mu^{(\varphi)} T^\mu  +  \xi_Q J_\mu^{(\varphi)}  Q^\mu  + \ldots + \mathcal{L}_\text{m}  \;,
		\end{split}
	\end{equation}
	where the matter Lagrangian $\mathcal{L}_\text{m}$ is only a function of $\varphi$ and $\xi$, $\xi_T$, $c_{TT}$,  $\tilde{c}_{TT}$, \etc are arbitrary real coefficients. Moreover, we defined
	\begin{equation} \label{realScalarSource}
		J_\mu^{(\varphi)} = \partial_\mu \varphi^2 \;.
	\end{equation}
	As before, the dots represent further analogous terms in which $T^\mu$ and/or $Q^\mu$ are replaced by some of the other tensors shown in \eqs \eqref{torsionTrace} to \eqref{torsionTensor} as well as \eqref{nonMetricityVector1} to \eqref{nonMetricityTensor}. Each of these additional contributions comes with an a priori undetermined coupling constant.
	
	In \eq \eqref{actionScalarFieldSchematic}, we have assumed as in \cite{Rigouzzo:2022yan}  that $\varphi$ obeys a discrete symmetry $\varphi \rightarrow - \varphi$. This condition is not essential and we impose it because the Higgs field of the SM (in unitary gauge) exhibits this property. Apart from the discrete symmetry, however, $\varphi$ represents at this point an arbitrary scalar field, not necessarily connected to the Higgs boson. Once matter is present, condition II.) matters. Without it, we could have included in addition to quadratic terms higher powers of $\varphi$. 
	Moreover, we remark that the non-minimal coupling to the Levi-Civita Ricci scalar,  $\varphi^2 \mathring{R}$, already exists in the metric formulation of GR. Loosely speaking, one could therefore say that \eq \eqref{actionScalarFieldSchematic} consists of all terms that are on the same footing as this non-minimal coupling to $ \mathring{R}$.
	
	The crucial novelty in \eq \eqref{actionScalarFieldSchematic} as compared to the case of pure gravity displayed in \eq \eqref{actionPureGravitySchematic} is the presence of terms linear in torsion or non-metricity, which couple to $J_\mu^{(\varphi)}$. These contributions act as source terms in the equations of motion and hence it follows that torsion and non-metricity no longer vanish. Schematically, we get
	\begin{equation} \label{solutionScalarFieldSchematic}
		T_\mu \sim \xi_T 	J_\mu^{(\varphi)} \;, \qquad Q_\mu \sim \xi_Q 	J_\mu^{(\varphi)}  \;,
	\end{equation}
	where again we restricted ourselves to exemplary contributions. We see that $T_\mu$ and $Q_\mu$ indeed feature one derivative, and so our derivative counting introduced in condition I.) is self-consistent.
	
	Plugging the solutions \eqref{solutionScalarFieldSchematic} back into our initial theory \eqref{actionScalarFieldSchematic}, we arrive at a Lagrangian of the form:
	\begin{equation} \label{actionScalarFieldSchematicEquivalent}
		\mathcal{L} \sim  \mathring{R} + \xi \varphi^2 \mathring{R} + f(\varphi^2) 	J_\mu^{(\varphi)} 	J^{(\varphi)\,\mu} + \mathcal{L}_\text{m}  \;.
	\end{equation}
	Here $f(\varphi)$ contains no derivatives, only depends on $\varphi^2$ and is determined by the various coupling constant that appear in \eq \eqref{actionScalarFieldSchematic}. Since solely the Levi-Civita connection appears in \eq \eqref{actionScalarFieldSchematicEquivalent}, we have derived an equivalent representation of our theory in the metric formulation of GR, in which the effects of torsion and non-metricity are replaced by a specific set of operators in the matter sector with mass dimension greater than four. It follows from conditions I.) and II.) that these new terms feature at most two derivatives, and hence they do not lead to any additional propagating degrees of freedom.
	Condition II.) is essential for the predictiveness of our approach. Without it, we could have included from the beginning arbitrary higher-dimensional operators in $\mathcal{L}_\text{m}$ and the specific subset of terms that are contained in $f(\varphi^2) 	J_\mu^{(\varphi)} 	J^{(\varphi)\,\mu}$ that arise due to torsion and non-metricity would not bring any new information. In other words, requirement II.) demands that \textit{non-renormalizable operators can only arise from coupling to gravity}. Of course, such a point of view can be motivated by the fact that GR is non-renormalizable from the outset. Needless to say, imposing conditions I.) and II.) amounts to an assumption and its viability remains to be determined. Arguably, the best way for doing so is to derive observable predictions and to compare them to measurements. 
	
	\paragraph*{Scalar field.} We have seen that a real scalar field $\varphi$ can couple to terms that are quadratic in torsion or non-metricity. Moreover, the term $J_\mu^{(\varphi)}$ as displayed in \eq \eqref{realScalarSource} interacts linearly with any of the four vectors $T^\mu$, $\hat{T}^\mu$, $Q^\mu$ and $\hat{Q}^\mu$ and hence sources torsion and non-metricity.
	Going beyond a real scalar field, we shall next discuss which other terms can source torsion or non-metricity, where we largely follow \cite{Karananas:2021zkl}. First, we consider a complex scalar field $\Phi$. If no other conditions are imposed, we can decompose it into two real scalar fields $\Phi_1$ and $\Phi_2$ as $\Phi = 1/\sqrt{2}(\Phi_1 + i \Phi_2)$. Then what we discussed above separately applies to $\Phi_1$ and $\Phi_2$. 
	
	More interesting is the case in which $\Phi$ is charged under a local $U(1)$-symmetry, as is \eg the case for the Higgs doublet in the SM. Then its absolute value,
	\begin{equation}
		|\Phi|^2 = \frac{1}{2} \varphi^2 \;,
	\end{equation} 
	allows for the same couplings as a real scalar field. On top of that, a second term can act as a source, namely the gauge-invariant current
	\begin{equation} \label{complexScalarSource}
		J_\mu^{(\Phi)} =	\frac{i}{2}\left(\Phi^\star \mathring{\mathcal{D}}_\mu \Phi - (\mathring{\mathcal{D}}_\mu \Phi)^\star \Phi \right) \;.
	\end{equation}
	Here $\mathring{\mathcal{D}}_\mu = \partial_\mu - i e \tilde{A}_\mu$ is the gauge-covariant derivative, where $e$ is the gauge coupling and $\tilde{A}_\mu$ represents the gauge field. Just like $J_\mu^{(\varphi)}$, the current $J_\mu^{(\Phi)}$ has mass dimension $3$ and features one derivative. Hence it can also couple linearly to the four vectors $T^\mu$, $\hat{T}^\mu$, $Q^\mu$ and $\hat{Q}^\mu$.
	
	\paragraph*{Fermion field.} Next, we consider a fermion field $\Psi$. Our discussion both applies to the case in which $\Psi$ is part of the SM and to fermions that are singlets under the gauge groups of the SM. Since we need to form bilinears, which by themselves already have mass dimension $3$, it is clear that fermions cannot couple to terms quadratic in torsion or non-metricity. However, we can form two source terms from the vector and axial currents:
	\begin{align} 
		J_\mu^{(V)} & = \bar{\Psi} \gamma_\mu \Psi \;, \label{vectorFermionSource}\\
		J_\mu^{(A)} & =  \bar{\Psi} \gamma^5 \gamma_\mu \Psi \;, \label{axialFermionSource}
	\end{align}
	where $\gamma_\mu$ and $\gamma^5$ correspond to gamma matrices. Just like the contributions discussed before, $J_\mu^{(V)}$ and $J_\mu^{(A)}$ carry one spacetime index and have mass dimension $3$. Therefore, requirements I.) and II.) imply that the interaction of the fermionic currents with torsion or non-metricity is fully analogous to the case of a scalar field, \ie $J_\mu^{(V)}$ and $J_\mu^{(A)}$ can couple linearly to any of the four vectors $T^\mu$, $\hat{T}^\mu$, $Q^\mu$ and $\hat{Q}^\mu$. However, a difference to the scalar sources exists since $J_\mu^{(V)}$ and $J_\mu^{(A)}$ do not contain a derivative. Correspondingly, the interaction of fermions with torsion and non-metricity leads to contributions of the form 
	\begin{equation}
		T_\mu \sim \bar{\Psi} \gamma_\mu \Psi + \ldots \;, \qquad Q_\mu \sim  \bar{\Psi} \gamma_\mu \Psi  +\ldots\;,
	\end{equation}
	which do not contain a derivative. In this sense, associating a derivative to torsion and non-metricity represents an assumption when it is exclusively sourced by fermions \cite{Diakonov:2011fs}.
	
	At first sight, a second category of terms seems to exist for coupling fermions to torsion or non-metricity. Namely, we can use the pure tensor parts $t_{\mu \nu \rho}$ and $q_{\mu \nu \rho}$ to construct the following terms:
	\begin{equation} \label{source_irrep}
		\bar{\Psi} \gamma^\mu \gamma^\nu \gamma^\rho \Psi t_{\mu \nu \rho}\;, \qquad  \bar{\Psi} \gamma^\mu \gamma^\nu \gamma^\rho \Psi q_{\mu \nu \rho} \;.
	\end{equation}
	However, using the property of the Dirac algebra, we have that  $\gamma^\mu \gamma^\nu \gamma^\rho=-g^{\mu \nu} \gamma^\rho-g^{\nu \rho} \gamma^\mu+g^{\mu \rho} \gamma^\nu+i \epsilon^{\sigma \mu \nu \rho} \gamma_\sigma \gamma^5$ and hence the two terms displayed in \eq \eqref{source_irrep} can be recast as (see also \cite{Karananas:2021zkl}\footnote{Note that there are sign differences as compared to \cite{Karananas:2021zkl}; see \eq \eqref{alpha_beta_mapping} and footnote \ref{fnSignDifference} further below.}):
	\begin{equation}
		\bar{\Psi} \gamma^\mu \gamma^\nu \gamma^\rho \Psi t_{\mu \nu \rho} =  -2 \bar{\Psi}\gamma^\alpha \Psi t^\mu_{~~\mu \alpha} +i \bar{\Psi} \gamma^\alpha \gamma^5 \Psi \epsilon^{\alpha \mu \nu \rho}t_{ \mu \nu \rho}\;,
	\end{equation}
	\begin{equation}
		\bar{\Psi} \gamma^\mu \gamma^\nu \gamma^\rho \Psi q_{\mu \nu \rho}= - \bar{\Psi} \gamma^\alpha \Psi q_{\alpha \mu}^{~~~\mu} \;.
	\end{equation}
	It is now clear that these terms vanish due to the properties of the pure tensor parts shown in \eqs \eqref{torsionTensor} and \eqref{nonMetricityTensor}. Therefore, we do not need to include them. 
	
	\paragraph*{Gauge field.} Finally, we discuss the effects of gauge fields, where we momentarily specialize to the case of $U(1)$. The coupling to the current \eqref{complexScalarSource} already leads to an interaction with the gauge field that is contained in the gauge covariant derivative. Due to the requirement of gauge invariance, the only other building block from which we can try to construct a coupling term is the field strength tensor $F_{\mu\nu}$ (see \eq \eqref{fieldStrength}). Since it already contains a derivative, it follows from condition I.) that it can only interact with a linear power of torsion or non-metricity. However, there is no torsion or non-metricity term with two spacetime indices. The same conclusion applies to gauge groups beyond $U(1)$, with the only difference that in this case an additional obstruction appears since the field strength tensor also carries an index of the gauge group. Thus, our requirements for constructing an action of matter coupled to metric-affine gravity do not allow for an independent coupling of gravity to gauge fields. 
	
	\paragraph*{Summary.}
	In summary, terms that are quadratic in torsion or non-metricity can only couple to a real scalar field $\varphi$ (or equivalently the absolute value of a complex scalar field) via
	\begin{equation}
		C = \varphi^2 \;.
	\end{equation}
	There are four contributions \eqref{realScalarSource}, \eqref{complexScalarSource}, \eqref{vectorFermionSource} and \eqref{axialFermionSource} for coupling matter linearly to torsion or non-metricity. The former two arise for scalar fields, where the second one only exists in the complex case, and the latter two are due to fermions. Thus, a source term generically reads 
	\begin{equation} \label{sourceGeneral}
		J_\mu = a^{(\varphi)} J_\mu^{(\varphi)} + a^{(\Phi)} J_\mu^{(\Phi)} +  a^{V} 	J_\mu^{(V)}   +  a^{A} 	J_\mu^{(A)} \;.
	\end{equation}
	Here $a^{(\varphi)}$, $ a^{(\Phi)}$, $a^V$ and $a^A$ are a priori undetermined coupling constants. There is no independent coupling between a gauge field and torsion or non-metricity. Moreover, no source term exists for the pure tensor parts $t_{\alpha \beta \gamma}$ and $q_{\alpha \beta \gamma}$.\footnote
	{See \cite{Wheeler:2023qyy} for a proposal to source the pure tensor parts from a field with spin $3/2$.} 
	For $J_\mu$ given in \eq \eqref{sourceGeneral}, we can solve for torsion and non-metricity and plug the result back into the action. In full analogy to \eq \eqref{actionScalarFieldSchematicEquivalent}, this leads to
	\begin{equation} \label{actionSchematicEquivalent}
		\mathcal{L} \sim  \mathring{R} + \xi \varphi^2 \mathring{R} + f(\varphi^2) 	J_\mu	J^{\mu} + \mathcal{L}_\text{m}  \;.
	\end{equation}
	As before, $f(\varphi^2)$, which is sensitive to all coupling constants related to torsion and non-metricity, only depends on $\varphi^2$ and contains no derivatives. It follows from \eq \eqref{sourceGeneral} that the contribution proportional to $J_\mu J^{\mu}$ generically contains ten different types of quadratic interactions among the different sources terms.
	
	\subsection{Relationship to previous works}
	In a very similar setting as ours, criteria for constructing an action of gravity coupled to matter were already proposed in \cite{Karananas:2021zkl}. In order to compare them to our requirements, we shall first repeat the conditions of \cite{Karananas:2021zkl}:
	\begin{enumerate}
		\item In the purely gravitational part of the action, only operators of mass dimension not greater than $2$ should appear.
		\item The matter Lagrangian action should only feature renormalizable operators in the flat limit, \ie for $g_{\mu\nu}= \eta_{\mu\nu}$ and $\Gamma_{~~\mu \lambda}^\alpha = 0$.
		\item Gravity and matter should only interact via operators of mass dimension not greater than $4$. 
	\end{enumerate}
	In the following, Arabic numerals shall refer to the criteria of \cite{Karananas:2021zkl} whereas Roman ones belong to our requirements listed section \ref{ssec:selectionrules}.
	
	On the one hand, our conditions I.) and II.) imply 1.) to 3.): Since in pure gravity the number of derivatives and the mass dimension of an operator coincide, requirement 1.) follows from I.). As in the absence of gravity non-renormalizability only arises from operators of mass dimension greater than $4$, criterion II.) implies 2.). Finally, it is evident that 3.) also follows from II.).
	
	On the other hand, our conditions I.) and II.) are slightly stronger than 1.) to 3.). In order to see this, we shall consider a $U(1)$-gauge field with field-strength tensor $F_{\mu\nu}$ (see \eq \eqref{fieldStrength}) and couple it to the pure tensor part $t^{\alpha \mu \nu}$ of torsion with a term
	\begin{equation} \label{tensorSource}
		\mathcal{L}_t = \mathring{\mathcal{D}}_\alpha F_{\mu\nu} t^{\alpha \mu \nu} \;,
	\end{equation}
	where $\mathring{\mathcal{D}}_\alpha$ is only covariant with respect to gravity.
	Since this contribution contains three derivatives, it is incompatible with our requirement I.). In contrast, it is admissible according to the criteria 1.) to 3.) of \cite{Karananas:2021zkl}.
	In the presence of the term \eqref{tensorSource}, $t_{\alpha \mu \nu}$ no longer vanishes but receives a contribution of the form
	\begin{equation} \label{sourcedPureTensor}
		t_{\alpha \mu \nu} \sim \mathring{\mathcal{D}}_\alpha F_{\mu\nu} + \frac{2}{3} g_{\alpha [\mu}  \mathring{\mathcal{D}}^\sigma F_{\nu]\sigma} \;,
	\end{equation}
	where we took into account that $t_{\alpha \mu \nu}$ only couples to the pure tensor part of $\mathring{\mathcal{D}}_\alpha F_{\mu\nu}$.\footnote
	{Correspondingly, $t_{\alpha \mu \nu}$ as shown in \eq \eqref{sourcedPureTensor} fulfills $g^{\alpha \nu} t_{\alpha \mu \nu} = 0$ and $\epsilon^{\alpha \beta \mu \nu} t_{\alpha \mu \nu} = 0$, in accordance with its definition \eqref{torsionTensor}.}
	After plugging the solution for torsion back into the action, a contribution with four derivatives of the form $\mathring{\mathcal{D}}_\alpha F_{\mu\nu} \mathring{\mathcal{D}}^\alpha F^{\mu\nu}$ will arise, among others. The term \eqref{tensorSource} is not the only contribution with more than two derivatives that would be permissible according to conditions 1.) to 3.). Other examples include $T_\mu A_\mu F^{\mu\nu}$ as well as
	$\mathring{\mathcal{D}}_\alpha \tilde{F}_{\mu\nu} t^{\alpha \mu \nu}$ with $\tilde{F}_{\mu\nu} = \epsilon_{\mu\nu\rho\sigma}F^{\rho\sigma}$ (see \cite{Pradisi:2022nmh}).

	Thus, conditions 1.) to 3.) would allow from some terms with more than two derivatives but only when an Abelian gauge field is involved. It is not clear why gauge fields should play a special role with regard to derivative counting. Moreover, criteria 1.) to 3.) would run counter to our guideline of finding an action that is as general as possible while excluding terms with more than two derivatives. Finally, the term \eqref{tensorSource} was not included in the action of EC gravity coupled to matter constructed in \cite{Karananas:2021zkl}, even though it conforms to the requirements 1.) to 3.) used there.
	Rather, employing our criteria I.) and II.) in the EC formulation of GR would lead to the theory considered in \cite{Karananas:2021zkl}.\footnote
	{Needless to say, this fact was part of our motivation for proposing conditions I.) and II.). We thank Georgios Karananas and Misha Shaposhnikov for discussions about this point.}
	Concerning our previous paper \cite{Rigouzzo:2022yan}, in which we studied the theory of a scalar field coupled to metric-affine gravity, we remark that our criteria and the conditions of \cite{Karananas:2021zkl} lead to identical results. Thus, the model of \cite{Rigouzzo:2022yan} can be derived equally well from our newly proposed requirements or the ones of \cite{Karananas:2021zkl}.

	\section{Details on computations of the equivalent metric theory}
	\label{app_details_computations}
		\subsection{Solving for torsion and non-metricity}
		\label{sapp_solving}
	Varying the action with respect to the six tensors $T^{\alpha}$, $\hat{T}^{\alpha}$, $t^{\alpha \beta \gamma}$, $Q^\gamma$, $\hat{Q}^\gamma$ and $q^{\alpha \beta \gamma}$, we obtain six equations of motions. For the action given in \eq \eqref{general_action_component}, they are: 
	\begin{equation}
		\begin{split}
			&2 C_2 \hat{T}^\alpha+C_3T^\alpha+E_2 Q^\alpha+E_4 \hat{Q}^\alpha=  J^\alpha_1 \;, \\
			&2 C_1 T^\alpha+C_3\hat{T}^\alpha+E_1Q^\alpha+E_3\hat{Q}^\alpha= J^\alpha_2 \;, \\
			&2 B_2\hat{Q}^\alpha+B_3Q^\alpha+E_3 T^\alpha+E_4\hat{T}^\alpha= J^\alpha_3 \;, \\
			&2B_1 Q^\alpha+B_3 \hat{Q}^\alpha+E_1 T^\alpha+E_2 \hat{T}^\alpha= J^\alpha_4 \;,
			\\&2 B_4 q_{\alpha \beta \gamma}+ 2B_5q_{(\beta \alpha) \gamma} +2D_2 \epsilon_{\alpha \lambda \delta (\beta}q^{\lambda \delta}_{~~\gamma )}+D_3 \epsilon_{\alpha \lambda \delta (\beta} t^{\lambda\delta}_{~~\gamma)}-E_5 t_{(\beta \gamma) \alpha}=0 \;,\\
			&2C_4 t_{\alpha \beta \gamma}+2D_1		\epsilon_{\alpha \lambda \delta [\beta}t^{\lambda \delta}_{~~\gamma ]}+D_3\epsilon_{\alpha \lambda \delta [\beta}q^{\lambda \delta}_{~~\gamma]}+E_5q_{[\beta \gamma]\alpha}=0 \;.
		\end{split}
	\end{equation}
	For the vector parts $T^\alpha$, $\hat{T}^\alpha$, $Q^\alpha$ and $\hat{Q}^\alpha$, the solutions are: 
	\begin{equation}
		T^\alpha=\frac{1}{Z}\sum_{i=1}^4 J_i^\alpha \mathcal{X}_i \;, \quad \hat{T}^\alpha=\frac{1}{Z}\sum_{i=1}^4 J_i^\alpha \mathcal{Y}_i \;, \quad	Q^\alpha=\frac{1}{Z}\sum_{i=1}^4 J_i^\alpha \mathcal{V}_i \;, \quad \hat{Q}^\alpha=\frac{1}{Z}\sum_{i=1}^4 J_i^\alpha \mathcal{W}_i	 \;.
		\label{solution_short}
	\end{equation}
	The common denominator reads
	\begin{equation}
		\begin{split} 
			Z&=	4 B_1 \left(B_2 (-4 C_1 C_2 + C_3^2) + C_1 E_4^2 + C_2 E_3^2 - C_3 E_3 E_4\right) \\
			&+ 4 B_2 C_1 E_2^2  + 4 B_2 C_2 E_1^2 - 4 B_2 C_3 E_1 E_2 + B_3^2 (4 C_1 C_2 - C_3^2)\\
			&+B_3 (- 4 C_1 E_2 E_4 -4 C_2 E_1 E_3  + 2 C_3 E_1 E_4 + 2 C_3 E_2 E_3)\\
			& - E_1^2 E_4^2  + 2 E_1 E_2 E_3 E_4 - E_2^2 E_3^2  \;,
		\end{split}
	\end{equation}
	and the numerators are: 
	\begin{equation}
		\begin{split}
			&\mathcal{X}_1=4 B_1 B_2 C_3 -2 B_1 E_3 E_4  -2 B_2 E_1 E_2 -B_3^2C_3+B_3 E_1 E_4 + B_3 E_2 E_3 \;, \\
			& \mathcal{X}_2=-8 B_1 B_2 C_2  +2  B_1 E_4^2 + 2 B_2 E_2^2 + 2 B_3^2 C_2-2 B_3 E_2 E_4 \;, \\
			& \mathcal{X}_3=4 B_1 C_2 E_3 -2 B_1 C_3 E_4  -2 B_3 C_2 E_1 + B_3 C_3 E_2 +E_1 E_2 E_4 - E_2^2 E_3 \;, \\
			& \mathcal{X}_4=4 B_2 C_2 E_1 -2 B_2 C_3 E_2 -2 B_3 C_2 E_3 +B_3 C_3 E_4  -E_1 E_4^2+E_2 E_3 E_4 \;,
		\end{split}
	\end{equation}
	\begin{equation}
		\begin{split}
			&\mathcal{Y}_1=-8 B_1 B_2 C_1 +2 B_1 E_3^2  +2 B_2 E_1^2 + 2 B_3^2 C_1-2 B_3 E_1 E_3  \;,\\
			& \mathcal{Y}_2=4 B_1 B_2 C_3 -2 B_1 E_3 E_4 -2 B_2 E_1 E_2 -B_3^2 C_3+B_3 E_2 E_3 +B_3 E_1 E_4 \;, \\
			& \mathcal{Y}_3=4  B_1 C_1 E_4 -2 B_1 C_3E_3 -2  B_3 C_1 E_2 + B_3 C_3E_1-E_1^2 E_4 +E_1 E_2 E_3 \;, \\
			& \mathcal{Y}_4=4  B_2 C_1 E_2 -2  B_2 C_3 E_1 -2 B_3 C_1 E_4 +B_3 C_3 E_3 +E_1 E_3 E_4 -E_2 E_3^2  \;,
		\end{split}
	\end{equation}
	\begin{equation}
		\begin{split}
			&\mathcal{V}_1=4 B_2 C_1 E_2 -2 B_2 C_3 E_1 -2  B_3 C_1 E_4 + B_3 C_3 E_3  +E_1  E_3 E_4  -E_2 E_3^2 \;, \\
			& \mathcal{V}_2=4 B_2 C_2 E_1 -2 B_2 C_3 E_2  -2 B_3 C_2 E_3 + B_3 C_3 E_4  -E_1 E_4^2+E_2 E_3 E_4  \;,\\
			& \mathcal{V}_3=4 B_3 C_1 C_2 -B_3 C_3^2 - 2 C_1 E_2 E_4 -2 C_2 E_1 E_3 +C_3 E_1 E_4 +C_3 E_2 E_3  \;,\\
			& \mathcal{V}_4=-8 B_2 C_1 C_2 + 2 B_2 C_3^2 +2 C_1 E_4^2 +2 C_2 E_3^2 -2  C_3 E_3 E_4  \;,
		\end{split}
	\end{equation}
	\begin{equation}
		\begin{split}
			&\mathcal{W}_1=4 B_1 C_1 E_4 -2  B_1 C_3 E_3-2  B_3 C_1 E_2 + B_3 C_3 E_1  - E_1^2 E_4+E_1 E_2 E_3 \;, \\
			& \mathcal{W}_2=4  B_1 C_2 E_3-2 B_1 C_3 E_4 -2 B_3 C_2 E_1 + B_3 C_3 E_2 +E_1 E_2 E_4 -E_2^2 E_3  \;,\\
			& \mathcal{W}_3=-8 B_1 C_1 C_2 + 2 B_1 C_3^2  +2 C_1 E_2^2 +2 C_2 E_1^2  -2  C_3 E_1 E_2 \;, \\
			& \mathcal{W}_4=4 B_3 C_1 C_2-B_3 C_3^2 -2 C_1 E_2 E_4 -2 C_2 E_1 E_3 +C_3 E_1 E_4 +C_3 E_2 E_3 \;. \\
		\end{split}
	\end{equation}
	Plugging back the solutions for torsion and non-metricity into the action \eqref{general_action_component}, we see that the source part \eqref{action_sources} simplifies to: 
	\begin{equation} \label{effective_action_sources}
		S_{\text{sources}}= \int \mathrm{d}^{4} x \sqrt{-g} \frac{1}{Z}\sum_{i=1}^4 \Big[- J_1^\alpha J_{i\alpha} \mathcal{Y}_i - J_2^\alpha J_{i\alpha} \mathcal{X}_i - J_3^\alpha J_{i\alpha} \mathcal{W}_i - J_4^\alpha J_{i\alpha} \mathcal{V}_i \Big] \;,
	\end{equation}
while the quadratic contribution \eqref{action_quadratic} becomes
	\begin{equation} \label{effective_action_quadratic}
		\begin{split}
			S_{\text{quadratic}}= &\int \mathrm{d}^{4} x \sqrt{-g} \frac{1}{Z^2}  \sum_{i,j=1}^4 J^\alpha_i J_{j \alpha}\Big[B_{1} \mathcal{V}_{i} \mathcal{V}_{j}+B_{2} \mathcal{W}_{i} \mathcal{W}_{j}+B_{3} \mathcal{V}_{i} \mathcal{W}_{j}\\
			&+C_{1}\mathcal{X}_{i} \mathcal{X}_{j}+C_{2} \mathcal{Y}_{i} \mathcal{Y}_{j}+C_{3} \mathcal{X}_{i} \mathcal{Y}_{j} \\
			&+E_1 \mathcal{X}_i \mathcal{V}_j+E_2 \mathcal{Y}_{i}\mathcal{V}_j+E_3 \mathcal{X}_{i} \mathcal{W}_j+E_4 \mathcal{Y}_i \mathcal{W}_j \Big] \;.
		\end{split}
	\end{equation}

	Now we have integrated out torsion and non-metricity, but the matter sector is still non-minimally coupled to gravity through the term $\Omega^2 \mathring{R}$ in $S_{\text{metric}}$ given (see \eq \eqref{action_metric}). To simplify the analysis, we go to the Einstein frame where gravity is minimally coupled to the matter sector. We achieve this by doing a conformal transformation of the metric: 
	\begin{equation}
		\begin{split}
			&g_{\alpha \beta} \rightarrow \Omega^{-2} g_{\alpha \beta} \;, \qquad g^{\alpha \beta} \rightarrow \Omega^{2} g^{\alpha \beta} \;,\\&
			e_\mu^{~A} \rightarrow \Omega^{-1}e_\mu^{~A} \;, \qquad	e^\mu_{~A} \rightarrow \Omega^{1}e^\mu_{~A} \;,\\
			& \sqrt{-g} \rightarrow \Omega^{-4} \sqrt{-g} \;, \\
			&g^{\alpha \beta} \mathring{R}_{\alpha \beta} \rightarrow \Omega^{2}[g^{\alpha \beta} \mathring{R}_{\alpha \beta}+6g^{\alpha \beta}(\mathring{\nabla}_\alpha \mathring{\nabla}_\beta \ln(\Omega)-\mathring{\nabla}_\alpha \ln(\Omega)\mathring{\nabla}_\beta \ln(\Omega))]
			\;.\\
		\end{split}
		\label{conformal_transfo}
	\end{equation}
	Torsion is not affected since it is completely independent of the metric. Note that technically, non-metricity transforms non-homogeneously under this conformal transformation, but at this point this is irrelevant since we have integrated out both torsion and non-metricity from the action. One can see from \eq \eqref{conformal_transfo} that  the scalar curvature $\mathring{R}$  transforms inhomogeneously due to the non-trivial dependence of the Levi-Civita connection $\mathring{\Gamma}^\gamma_{~\alpha \beta}$ on the metric $g_{\mu\nu}$. This leads to a modification of the kinetic terms of the matter sector, depending on the function $\Omega$. After applying criterion II, we shall see that $\Omega$ can only be a function of the norm $\varphi$ of the complex scalar field $\Phi$. Therefore only the kinetic term of the complex scalar field will receive contribution from the inhomogeneous part of the transformation of the scalar curvature. After performing the conformal transformation  \eqref{conformal_transfo}, the metric action becomes: 
	\begin{equation} \label{action_metric_after_conformal}
		\begin{split}
			S_{\text{metric}}= &\int \mathrm{d}^{4} x \sqrt{-g} \Big[\frac{1}{2}  \mathring{R}-\frac{V(\Phi,\Psi, F_{\mu \nu})}{\Omega^4} - \frac{3}{\Omega^2}g^{\alpha \beta} \mathring{D}_{\alpha}\Omega \mathring{D}_{\beta}\Omega\\
			&  - \frac{\tilde{K}_1}{\Omega^2} g^{\alpha \beta} \mathring{D}_{\alpha} \Phi \mathring{D}_{\beta}  \Phi^\star+\frac{i\tilde{K}_2}{2 \Omega^3} ( \bar{\Psi}\gamma^\mu \mathring{D}_\mu \Psi- \overline{\mathring{D}_\mu \Psi} \gamma^\mu \Psi)-\frac{1}{4}\tilde{K}_3g^{\mu \alpha} g^{\nu \beta}F_{\mu \nu} F_{\alpha \beta} \Big] \;.
		\end{split}
	\end{equation}
	We recall that the gamma matrices come with a tetrad $\gamma_\mu=e_\mu^{~A} \gamma_{A}$, so this is why the kinetic term for the fermionic field in \eq \eqref{action_metric_after_conformal} gets shifted by a factor of $\Omega^{-3}$, whereas the kinetic term for the complex scalar field is only shifted by $\Omega^{-2}$. We note that the kinetic term for the vector field does not change.
	On the other hand, the sources and quadratic part of the action are simply multiplied by a factor of $\Omega^{-2}$, coming from the combination of the determinant of the metric and a hidden metric tensor there to contract indices. 
	
	 Finally, we can write suggestively the sum of these two actions as: 
	\begin{equation} \label{effective_action_sum}
		\begin{split}
			S_{\text{sources}}+S_{\text{quadratic}}= &\int \mathrm{d}^{4} x  \dfrac{\sqrt{-g}}{\Omega^2}\Big[ (J_1^\alpha)^2 \mathcal{L}_{11}+ (J_2^\alpha)^2 \mathcal{L}_{22}+(J_3^\alpha)^2 \mathcal{L}_{33}+(J_4^\alpha)^2 \mathcal{L}_{44} \\
			& +J_1^\alpha J_{2 \alpha} \mathcal{L}_{12}+ J_1^\alpha J_{3 \alpha} \mathcal{L}_{13}+ J_1^\alpha J_{4 \alpha} \mathcal{L}_{14} \\ 
			& +J_2^\alpha J_{3 \alpha} \mathcal{L}_{23}+ J_2^\alpha J_{4 \alpha} \mathcal{L}_{24} + J_3^\alpha J_{4 \alpha} \mathcal{L}_{34}\Big] \;,
		\end{split}
	\end{equation}
	where the $\mathcal{L}_{ij}$ are coefficients that depend only on the other parameters.\footnote{More explicitly, $\mathcal{L}_{ij}$ solely depends on the remaining coefficients of interest, \ie $B_1$, $B_2$, $B_3$, $C_1$, $C_2$, $C_3$, $E_1$, $E_2$, $E_3$ and $E_4$.} We can express them compactly as: 
	\begin{equation} \label{L_ij}
		\begin{split}
			& \mathcal{L}_{11}=-\frac{\mathcal{Y}_1}{Z}+\frac{\mathcal{Q}_{11}}{Z^2} \;, \quad \mathcal{L}_{22}=-\frac{\mathcal{X}_2}{Z}+\frac{\mathcal{Q}_{22}}{Z^2} \;, \quad \mathcal{L}_{33}=-\frac{\mathcal{W}_3}{Z}+\frac{\mathcal{Q}_{33}}{Z^2} \;, \quad \mathcal{L}_{44}=-\frac{\mathcal{V}_4}{Z}+\frac{\mathcal{Q}_{44}}{Z^2} \;, \\
			& \mathcal{L}_{12}=-\frac{\mathcal{Y}_2+\mathcal{X}_1}{Z}+\frac{2\mathcal{Q}_{12}}{Z^2} \;, \quad \mathcal{L}_{13}=-\frac{\mathcal{Y}_3+\mathcal{W}_1}{Z}+\frac{2\mathcal{Q}_{13}}{Z^2} \;, \quad \mathcal{L}_{14}=-\frac{\mathcal{Y}_4+\mathcal{V}_1}{Z}+\frac{2\mathcal{Q}_{14}}{Z^2}\;, \\ 
			&\mathcal{L}_{23}=-\frac{\mathcal{W}_2+\mathcal{X}_3}{Z}+\frac{2\mathcal{Q}_{23}}{Z^2} \;, \quad \mathcal{L}_{24}=-\frac{\mathcal{V}_2+\mathcal{X}_4}{Z}+\frac{2\mathcal{Q}_{24}}{Z^2}  \;, \quad \mathcal{L}_{34}=-\frac{\mathcal{W}_4+\mathcal{V}_3}{Z}+\frac{2\mathcal{Q}_{34}}{Z^2} \;,
		\end{split}
	\end{equation}
	where the coefficients $\mathcal{Q}_{ij}$ are given by:
	\begin{equation}
		\begin{split}
			\mathcal{Q}_{ij} = 
			& B_1 \mathcal{V}_{i}\mathcal{V}_{j}+ B_2 \mathcal{W}_{i}\mathcal{W}_{j}+ B_3 \mathcal{V}_{(i}\mathcal{W}_{j)} +  C_1 \mathcal{X}_{i}\mathcal{X}_{j}+ C_2 \mathcal{Y}_{i}\mathcal{Y}_{j}\\ &+ C_3 \mathcal{X}_{(i}\mathcal{Y}_{j)} + E_1 \mathcal{X}_{(i} \mathcal{V}_{j)}  + E_2 \mathcal{Y}_{(i} \mathcal{V}_{j)}+ E_3 \mathcal{X}_{(i} \mathcal{W}_{j)}  + E_4 \mathcal{Y}_{(i} \mathcal{W}_{j)} \;.
		\end{split}
	\end{equation}
As a result, we obtain \eq \eqref{final_lagrangian} shown in the main part.
	
		\subsection{Explicit interaction terms}
		\label{sapp:interaction_terms}
Next we perform a field redefinition of the fermionic field $\Psi$ to canonically normalize its kinetic term: 
	\begin{equation}
		\Psi \rightarrow \Omega^{3/2} \Psi .
	\end{equation}
	After this second conformal transformation of the fermionic field and also taking into account \eq \eqref{selec_rulE_3}, the action reads:
	\begin{equation} \label{action_metric_after_conformal_2}
		\begin{split}
			S_{\text{metric}}= &\int \mathrm{d}^{4} x \sqrt{-g} \Big[\frac{1}{2}  \mathring{R}-\frac{V(\Phi,\Omega^{3/2}\Psi, F_{\mu \nu})}{(1+\xi \varphi^2)} - \frac{3\xi^2}{4(1+\xi \varphi^2)^2}g^{\alpha \beta} \mathring{D}_{\alpha}\varphi^2 \mathring{D}_{\beta}\varphi^2\\
			&  - \frac{1}{1+\xi \varphi^2} g^{\alpha \beta} \mathring{D}_{\alpha} \Phi \mathring{D}_{\beta}  \Phi^\star+\frac{i}{2
			} ( \bar{\Psi}\gamma^\mu \mathring{D}_\mu \Psi- \overline{\mathring{D}_\mu \Psi} \gamma^\mu \Psi)-\frac{1}{4}g^{\mu \alpha} g^{\nu \beta}F_{\mu \nu} F_{\alpha \beta} \Big] \;.
		\end{split}
	\end{equation}
	Clearly the potential term $V$ may change after this second transformation and the fermionic currents defined in \eq \eqref{new_currents_2} and \eq \eqref{new_currents_3} will also be shifted to: 
	\begin{equation}
		V_\alpha \rightarrow \Omega^2 V_\alpha \;, \qquad A_\alpha \rightarrow \Omega^2 A_\alpha  \;.
	\end{equation}
	Substituting the $4$-current $J^\alpha_i$ given by \eq \eqref{selec_rulE_3} into \eq \eqref{effective_action_sum}, we arrive at interactions of the form $V$-$V$, $V$-$A$, $V$-$S$, etc which can be written compactly as: 
	\begin{align} \label{eq1}
		\mathcal{L}_{AA} &= \Omega^{2} (\sum_{i\leq j}^{4} \mathcal{L}_{ij}a^A_i a^A_j)A_{\mu}A^\mu \;,\\ \label{eq2}
		\mathcal{L}_{VV} &= \Omega^{2}(\sum_{i\leq j}^{4} \mathcal{L}_{ij}a^V_i a^V_j)V_{\mu}V^\mu \;,\\
		\mathcal{L}_{\varphi \varphi} &= \Omega^{-2}(\sum_{i\leq j}^{4} \mathcal{L}_{ij}a^{(\varphi)}_i a^{(\varphi)}_j) \mathring{D}^\mu \varphi^2\mathring{D}_\mu \varphi^2\;,\\
		\mathcal{L}_{\Phi \Phi}&=\Omega^{-2}(\sum_{i\leq j}^4 \mathcal{L}_{ij}a^{(\Phi)}_i a^{(\Phi)}_j) S^\mu S_\mu  \;,\\
		\mathcal{L}_{AV} &= 2\Omega^{2}(\sum_{i\leq j}^{4} \mathcal{L}_{ij}a^A_{(i} a^V_{j)})A_{\mu}V^\mu \;,\\
		\mathcal{L}_{A \varphi} &=2 (\sum_{i\leq j}^{4} \mathcal{L}_{ij}a^A_{(i} a^{(\varphi)}_{j)})A_{\mu}\mathring{D}^\mu \varphi^2 \;,\\
		\mathcal{L}_{A \Phi} &=2(\sum_{i\leq j}^{4} \mathcal{L}_{ij}a^A_{(i} a^{(\Phi)}_{j)})A_{\mu}S^\mu \;,\\
		\mathcal{L}_{V \varphi} &= 2(\sum_{i\leq j}^{4} \mathcal{L}_{ij}a^V_{(i} a^{(\varphi)}_{j)})V_{\mu}\mathring{D}^\mu \varphi^2 \;,\\
		\mathcal{L}_{V \Phi} &=2 (\sum_{i\leq j}^{4} \mathcal{L}_{ij}a^V_{(i} a^{(\Phi)}_{j)})V_{\mu}S^\mu \;,\\ \label{eq10}
		\mathcal{L}_{\varphi \Phi} &=2 \Omega^{-2}(\sum_{i\leq j}^{4} \mathcal{L}_{ij}a^{(\Phi)}_{(i} a^{(\varphi)}_{j)})S_{\mu}\mathring{D}^\mu \varphi^2 \;.
	\end{align}
	Notice that all the buildings blocks we need are the $\mathcal{L}_{ij}$ coefficients, which we have derived previously. The sum on $i$,$j$ consists generally of $10$ terms, because the whole expression is symmetric in $i$,$j$. All types of interactions are spawn, and the form of equations is identical up to that point. The only difference comes from the power of the $\Omega$ factors. Indeed, because of the conformal transformation of the metric and of the fermion field, each term acquires a different power of $\Omega$ depending on the presence of fermionic current. From \eqs \eqref{eq1} to \eqref{eq10}, we can derive \eqs \eqref{eqAA} to \eqref{eqSh} in the main part.

		\subsection{Known limits}
	\label{sapp:knownlimits}
	Finally, we shall show that our result \eqref{final_final_lagrangian} reproduces the previous findings of \cite{Karananas:2021zkl} and \cite{Rigouzzo:2022yan} in specific limits.
	\paragraph{Reproducing \cite{Rigouzzo:2022yan}:}
	In \cite{Rigouzzo:2022yan}, the only matter field present is a real scalar field. Therefore to reproduce previous results we need to set the kinetic terms for the fermion and the vector field to zero in \eq \eqref{final_lagrangian}. We also take the limit where $a_j^{(\Phi)}$, $a_j^{A}$ and $a_j^V$ vanish in \eq \eqref{selec_rulE_3}. Consequently, only $\mathcal{L}_{\varphi \varphi}$ as displayed in \eq \eqref{eqhh} will be left:
	\begin{equation}
		\mathcal{L}_{\varphi \varphi} =\Omega^{-2} \frac{\sum_{n=0}^{3} P^{(\varphi \varphi)}_n \varphi^{2 n}}{D} \mathring{D}^\mu \varphi^2\mathring{D}_\mu \varphi^2 \; .
	\end{equation}
	The action will be much simpler and given by: 
	\begin{equation}
		S=\int \mathrm{d}^4 x \sqrt{-g} \left[\frac{1}{2} \mathring{R}-\frac{1}{2} K(\varphi) g^{\alpha \beta} \partial_\alpha \varphi \partial_\beta \varphi-\frac{V(\varphi)}{\Omega^4}\right] \;.
	\end{equation}
	Gathering together all contributions to the kinetic term of the scalar field, we obtain an expression for the modified kinetic term: 
	\begin{equation}
		K(\varphi)= \frac{1}{1+\xi \varphi^2}\left(1+\frac{6\varphi^2\xi^2}{1+\xi h^2}-8\varphi^2\frac{\sum_{n=0}^3 P^{(\varphi \varphi)}_n \varphi^{2 n}}{\sum_{m=0}^4 O_m \varphi^{2 m}}\right) \;,
	\end{equation}
	which exactly reproduces \eq (3.16) of \cite{Rigouzzo:2022yan} if we relabel $\varphi \rightarrow h$ and absorb the factor of $8$ into the polynomials $P^{(\varphi \varphi)}_n$.
	\paragraph{Reproducing \cite{Karananas:2021zkl}:} In \cite{Karananas:2021zkl}, all matter fields are present. However, the EC formulation of GR is employed, \ie non-metricity is assumed to be absent. Therefore, we need to set $J_3^\alpha$, $J_4^\alpha$, all $B_i$'s and $E_i's$ coefficients to zero. This simplifies greatly the form of the solution for torsion: 
	\begin{equation}
		T^\mu = \frac{-C_3 J^\mu_1+2C_2 J^\mu_2}{Z} \;, \qquad \hat{T}^\mu= \frac{-C_3 J^\mu_2+2C_1 J^\mu_1}{Z} \;,
	\end{equation}
	with the common denominator: 
	\begin{equation}
		Z= 4 C_1C_2 - C_3^2 \; .
	\end{equation}
	Only $\mathcal{L}_{11}$, $\mathcal{L}_{12}$ and $\mathcal{L}_{22}$ will contribute: 
	\begin{equation}
		S_{\text{sources}}+S_{\text{quadratic}}= \int \mathrm{d}^{4} x  \dfrac{\sqrt{-g}}{\Omega^2}\Big[ (J_1^\alpha)^2 \mathcal{L}_{11}+ (J_2^\alpha)^2 \mathcal{L}_{22} +J_1^\alpha J_{2 \alpha} \mathcal{L}_{12}\Big] \;,
	\end{equation}
	with the $\mathcal{L}_{ij}$ being much simpler: 
	\begin{equation}
		\mathcal{L}_{11}=-\frac{C_1}{4 C_1 C_2-C_3^2} \;, \quad \mathcal{L}_{12}=\frac{C_3}{4 C_1 C_2-C_3^2} \;, \quad \mathcal{L}_{22}=-\frac{C_2}{4 C_1 C_2-C_3^2} \;.
	\end{equation}
	The corresponding torsion-induced interactions are: 
	\begin{align}
		\mathcal{L}_{AA}&=\Omega^2 \frac{C_1 (a_1^{A})^2+C_2 (a_2^{A})^2-C_3 a_1^A a_2^A}{C_3^2-4 C_1 C_2} A_\mu A^\mu \\
		\mathcal{L}_{VV} &= \Omega^2  \frac{C_1 (a_1^{V})^2+C_2 (a_2^{V})^2-C_3 a_1^V a_2^V}{C_3^2-4 C_1 C_2}V_{\mu}V^\mu \;,\\
		\mathcal{L}_{\varphi \varphi} &=\Omega^{-2} \frac{C_1 (a_1^{(\varphi)})^2+C_2 (a_2^{(\varphi)})^2-C_3 a_1^{(\varphi)} a_2^{(\varphi)}}{C_3^2-4 C_1 C_2} \mathring{D}^\mu \varphi^2\mathring{D}_\mu \varphi^2 \;,\\
		\mathcal{L}_{\Phi \Phi}&=\Omega^{-2} \frac{C_1 (a_1^{(\Phi)})^2+C_2 (a_2^{(\Phi)})^2-C_3 a_1^{(\Phi)} a_2^{(\Phi)}}{C_3^2-4 C_1 C_2}S^\mu S_\mu  \;,\\
		\mathcal{L}_{AV} &= \Omega^{2}\frac{2C_1 a_1^Aa_1^V+2C_2 a_2^{A}a_2^V -C_3 (a_2^A a_1^V+a_1^A a_2^V)}{ C_3^2-4 C_1 C_2}A_{\mu}V^\mu \;,\\
		\mathcal{L}_{A \varphi} &= \frac{2C_1 a_1^A a_1^{(\varphi)}+2C_2 a_2^{A}a_2^{(\varphi)} -C_3 (a_2^A a_1^{(\varphi)}+a_1^A a_2^{(\varphi)})}{C_3^2-4 C_1 C_2}A_{\mu}\mathring{D}^\mu \varphi^2 \;,\\
		\mathcal{L}_{A \Phi} &=  \frac{2C_1 a_1^A a_1^{(\Phi)}+2C_2 a_2^{A}a_2^{(\Phi)} -C_3 (a_2^A a_1^{(\Phi)}+a_1^A a_2^{(\Phi)})}{C_3^2-4 C_1 C_2}A_{\mu}S^\mu \;,\\
		\mathcal{L}_{V \varphi} &=  \frac{2C_1 a_1^V a_1^{(\varphi)}+2C_2 a_2^{V}a_2^{(\varphi)} -C_3 (a_2^V a_1^{(\varphi)}+a_1^V a_2^{(\varphi)})}{C_3^2-4 C_1 C_2}V_{\mu}\mathring{D}^\mu \varphi^2 \;,\\
		\mathcal{L}_{V \Phi} &=  \frac{2C_1 a_1^V a_1^{(\Phi)}+2C_2 a_2^{V}a_2^{(\Phi)} -C_3 (a_2^V a_1^{(\Phi)}+a_1^V a_2^{(\Phi)})}{C_3^2-4 C_1 C_2}V_{\mu}S^\mu \;,\\
		\mathcal{L}_{\varphi \Phi} &= \Omega^{-2} \frac{2C_1 a_1^{(\varphi)} a_1^{(\Phi)}+2C_2 a_2^{(\varphi)}a_2^{(\Phi)} -C_3 (a_2^{(\varphi)} a_1^{(\Phi)}+a_1^{(\varphi)} a_2^{(\Phi)})}{C_3^2-4 C_1 C_2}S_{\mu}\mathring{D}^\mu \varphi^2 \;.
	\end{align}
	This exactly reproduces the main results of \cite{Karananas:2021zkl}, equations (52) to (61), where we show in table \ref{dictionary} the correspondence between the different notations.
	\begin{table}[H]
		\centering
		\begin{tabular}{|c|c|c|c|c|c|c|c|c|c|}
			\hline
			Current paper & $T^\mu$ & $\hat{T}^\mu$ & $C_1$ & $C_2$ & $C_3$ & $a_{1/2}^V$ & $a_{1/2}^A$ & $a_{1/2}^{(\Phi)}$ & $a_{1/2}^{(\varphi)} D_\mu \varphi^2$ \\ \hline
			Paper \cite{Karananas:2021zkl} & $v^\mu$ & $a^\mu$ & $\frac{G_{vv}}{2}$ & $\frac{G_{a a}}{2}$ & $G_{a v}$ & $-\zeta^{a /v}_{V}$ & $-\zeta^{ a /v}_{A}$ & $-Z^{a/v}_{S}$ & $-\partial_\mu Z^{a/v}_{\Phi}$ \\ \hline
		\end{tabular}
		\caption{Correspondence of notations between the present analysis and \cite{Karananas:2021zkl}.}
		\label{dictionary}
	\end{table}

	\subsection{Portal to dark matter in metric-affine gravity}
	\label{subsec:portaltoDM}
The polynomials in \eqs \eqref{FourFermionMAG} and \eqref{FourFermionMAG2} are given as follows. The common denominator reads
	\begin{equation}
		\begin{split}
			O_0=& 4 b_{30} c_{20} e_{10} e_{30}-2 b_{30} c_{30} e_{20} e_{30}-2 b_{30} c_{30} e_{10} e_{40}+4 b_{30} c_{10} e_{20} e_{40}-4 b_{20} c_{20} e_{10}^2-4 b_{20} c_{10} e_{20}^2
			\\&-4 b_{10} c_{20} e_{30}^2-4 b_{10} c_{10} e_{40}^2+4 b_{20} c_{30} e_{10} e_{20}+4 b_{10} c_{30} e_{30} e_{40}+b_{30}^2 c_{30}^2-4 b_{30}^2 c_{10} c_{20}\\&
			-4 b_{10} b_{20} c_{30}^2  +16 b_{10} b_{20} c_{10} c_{20}+e_{20}^2 e_{30}^2+e_{10}^2 e_{40}^2-2 e_{10} e_{20} e_{30} e_{40} \;,
		\end{split}
	\end{equation}
	and the polynomials in the numerators all have the same form: 
	\begin{equation}
		\begin{split}
			P_0^{(\circ \bullet)}&=  a_1^{(\circ} a_1^{\bullet)} \left(b_{30}^2 c_{10}-4 b_{10} b_{20} c_{10}-b_{30} e_{10} e_{30}+b_{20} e_{10}^2+b_{10} e_{30}^2\right)
			\\ &+a_2^{(\circ} a_2^{\bullet)} \left(b_{30}^2 c_{20}-4 b_{10} b_{20} c_{20}-b_{30} e_{20} e_{40}+b_{20} e_{20}^2+b_{10} e_{40}^2\right)
			\\&+a_3^{(\circ} a_3^{\bullet)} \left(b_{10} c_{30}^2-4 b_{10} c_{10} c_{20}-c_{30} e_{10} e_{20}+c_{20} e_{10}^2+c_{10} e_{20}^2\right)
			\\ & +a_4^{(\circ} a_4^{\bullet)} \left(b_{20} c_{30}^2-4 b_{20} c_{10} c_{20}-c_{30} e_{30} e_{40}+c_{20} e_{30}^2+c_{10} e_{40}^2\right)
			\\ & + a_1^{(\circ}a_2^{\bullet)} \left(-b_{30}^2 c_{30}+4 b_{10} b_{20} c_{30}+b_{30} e_{20} e_{30}+b_{30} e_{10} e_{40}-2 b_{20} e_{10} e_{20}-2 b_{10} e_{30} e_{40}\right)
			\\& +a_3^{(\circ}a_4^{\bullet)} \left(-b_{30} c_{30}^2+4 b_{30} c_{10} c_{20}+c_{30} e_{20} e_{30}+c_{30} e_{10} e_{40}-2 c_{20} e_{10} e_{30}-2 c_{10} e_{20} e_{40}\right)
			\\ &+a_1^{(\circ} a_3^{\bullet)} \left(b_{30} c_{30} e_{10}-2 b_{30} c_{10} e_{20}-2 b_{10} c_{30} e_{30}+4 b_{10} c_{10} e_{40}-e_{40} e_{10}^2+e_{20} e_{30} e_{10}\right)
			\\& +a_1^{(\circ} a_4^{\bullet)} \left(b_{30} c_{30} e_{30}-2 b_{20} c_{30} e_{10}+4 b_{20} c_{10} e_{20}-2 b_{30} c_{10} e_{40}-e_{20} e_{30}^2+e_{10} e_{40} e_{30}\right)
			\\ &+a_2^{(\circ} a_3^{\bullet)} \left(b_{30} c_{30} e_{20}-2 b_{30} c_{20} e_{10}+4 b_{10} c_{20} e_{30}-2 b_{10} c_{30} e_{40}-e_{30} e_{20}^2+e_{10} e_{40} e_{20}\right)
			\\ &+a_2^{(\circ} a_4^{\bullet)}\left(b_{30} c_{30} e_{40}+4 b_{20} c_{20} e_{10}-2 b_{20} c_{30} e_{20}-2 b_{30} c_{20} e_{30}-e_{10} e_{40}^2+e_{20} e_{30} e_{40}\right)	 \;,	
		\end{split}
	\end{equation}
	where $\circ$ and $\bullet$ can take the value $V$, $A$, $\varphi$.

	\bibliographystyle{utphys}
	\bibliography{EC}

\end{document}